\documentclass[10pt,a4paper,twocolumn,aps,pra,showpacs,noshowkeys,superscriptaddress,longbibliography]{revtex4-2}

\usepackage[T1]{fontenc}
\usepackage[latin9]{inputenc}
\usepackage{color}
\usepackage{amsmath}
\usepackage{amssymb}
\usepackage{graphicx}
\usepackage[breaklinks=true,colorlinks=true,citecolor=blue,filecolor=black,linkcolor=red,urlcolor=blue]{hyperref}
\usepackage[normalem]{ulem}

\begin{document}
\title{Nonclassical Properties and Anderson Localization of Quantum States in Coupled Waveguides}

\author{Thais L. Silva}
\affiliation{Instituto de Física, Universidade Federal de Goiás, 74.690-900, Goiânia,
Goiás, Brazil}
\affiliation{Instituto de Física,  Universidade Federal do Rio de Janeiro, 21941-972, Rio de Janeiro, RJ, Brazil}
\author{Wesley B. Cardoso}
\affiliation{Instituto de Física, Universidade Federal de Goiás, 74.690-900, Goiânia,
Goiás, Brazil}
\author{Ardiley T. Avelar}
\affiliation{Instituto de Física, Universidade Federal de Goiás, 74.690-900, Goiânia,
Goiás, Brazil}
\author{Jorge M. C. Malbouisson}
\affiliation{Instituto de Física, Universidade Federal da Bahia, 40.210-340, Salvador,
Bahia, Brazil}

\begin{abstract}
{We consider the propagation of light beams through  disordered lattices of coupled waveguides searching for Anderson localization and investigating the evolution of nonclassical properties of injected quantum states. We assume that the beam is initially in a variety of states, such as the complementary coherent state, the reciprocal binomial state and the polynomial state. The statistical properties of the evolved states were analyzed numerically as functions of the localization/delocalization parameters averaged over many realizations of disorder. We also numerically reconstruct the Wigner function of the output state. Interestingly, we find that high values of disorder tend to preserve quantum properties of some input states when we look at the input waveguide despite of the coupling between it and the neighboring waveguides.
}
\end{abstract}

\maketitle

\section{Introduction}

\quad Nonclassical properties of quantized light field
and its generation {--} i.e., the \textquotedblleft quantum states engineering\textquotedblright{--} 
{are} essential ingredient{s} of quantum optics and quantum information.
Indeed, nonclassical states perform a crucial role in many potential
applications such as quantum teleportation \cite{Bennett_PRL93,Bouwmeester_NAT97,Boschi_PRL98},
quantum cryptography \cite{Bennett_JC92}, quantum computation and
quantum communication \cite{Nielsen_10}, quantum internet \cite{Kimble_NAT08},
etc.  
{In this direction}, periodic photonic systems {has} emerge{d} as a platform to engineer
new light field structures, presenting numerous significant technological
advances \cite{Joannopoulos_11,Joannopoulos_NAT97,Inoue_04,Johnson_01}.
On the other hand, the study of nonperiodic photonic structures, by
using small defects in periodic lattices \cite{Ishizaki_NP13,Rinne_NP08,Braun_AM06}
or disordered and quasi-periodic structures \cite{Wiersma_NAT97,Schwartz_NAT07,Lahini_PRL08,Garcia_PRA11,Sapienza_SCI10},
has demonstrated a diversity of optical effects in the presence of
Anderson Localization, such as engineering of random lasers \cite{Liu_NN14}
and structurally colored materials with precisely controllable
wavelength and angular dependence of scattering \cite{Rockstuhl_13}.
Recently, it was demonstrated how to tune and freeze disorder in
photonic crystals by using percolation lithography \cite{Burgess_SR16}.

Anderson localization -- i.e., the suppression of transport due
to destructive interference of the many paths associated with coherent
multiple scattering from the modulation of a disordered potential
\cite{Anderson_PR58}-- has been experimentally observed in microwaves
\cite{Chabanov_NAT00}, light diffusive media \cite{Storzer_PRL06,Wiersma_NAT97},
photonic crystals \cite{Schwartz_NAT07,Lahini_PRL08}, Bose-Einstein
condensates \cite{Billy_NAT08,Roati_NAT08}, sound waves \cite{Hu_NP08},
optical fiber arrays \cite{Srinivasan_PRA08}, etc. Inspired by these
experimental investigations, many theoretical studies have been proposed
by considering the system in the presence of disordered potentials
(see for example \cite{Cheng_PRA10,Muruganandam_JPB10,Cheng_PRA11,Cheng_PRA11-2,Cheng_PRA11-3,Cardoso_NA12,Cheng_PRA14,Xi_PB15,Cardoso_OQE16,Cardoso_PLA19,Santos_ND20}).

Recently, Anderson localization of nonclassical light was investigated
for propagation in an array of waveguides in which neighboring waveguides
are evanescently coupled and disorder can be added in
a controlled manner \cite{Thompson_PRA10}. Specifically, 
{that} work 
{investigated} the consequences of using both sub-Poissonian and
super-Poissonian input light on the characteristics of Anderson localization, 
{verifying} the enhancement in fluctuations of localized light and superbunching
due to the medium\textquoteright s disorder.  Also,  an
important consequence of sub-Poissonian statistics of the incoming
light is to quench the total fluctuations at the output \cite{Thompson_PRA10}.
The system employed in Ref.  \cite{Thompson_PRA10} is similar to
that used in Ref.  \cite{Lahini_PRL08} to experimentally investigate
the evolution of linear and nonlinear waves  {in the presence} of 
Anderson {localization}.  {Moreover, the disordered one-dimensional waveguide lattice
was also used to experimentally investigate an extensive list of phenomena:} the signature of a localization
phase transition for light by directly measuring wave transport inside
the lattice \cite{Lahini_PRL09}, quantum correlations between noninteracting
particles evolving simultaneously in a disordered medium \cite{Lahini_PRL10},
Hanbury Brown and Twiss correlations of Anderson localized waves \cite{Lahini_PRA11},
the control of the polarization state of coherent light propagating
through an optically thick multiple scattering medium by controlling
only the spatial phase of the incoming field with a spatial light
modulator \cite{Guan_OL12}, the coherent manipulation of two-photon
path-entangled states by multimode interference in multimode waveguides
\cite{Poem_PRL12}, the observation of topological phase transitions
in photonic quasicrystals \cite{Verbin_PRL13}, the observation of
ensemble-averaged quantum correlations between path-entangled photons
undergoing Anderson localization \cite{Gilead_PRL15}, the two-photon
Anderson localization in a quadratic waveguide array with the emergence
of off-diagonal disorder \cite{Bai_JO16}, etc.

Here, inspired by the results obtained in Ref. \cite{Thompson_PRA10},
we numerically investigate the propagation of light {beams,} previously
prepared in nonclassical states of the electromagnetic field{,} propagating
in disordered lattice{s of waveguides} and undergoing Anderson localization. Our
goal is to verify the influence of Anderson localization on the statistical
properties of previously prepared input light field (nonclassical
states). To this end, we assume the beam in a variety of states, {namely,}
the complementary coherent state \cite{Avelar_JOB04}, reciprocal
binomial state \cite{Valverde_PLA03}, polynomial state \cite{Souza_OC04},
thermal state, coherent state, and squeezed state \cite{Walls_08}.

The rest of the paper is organized as follows. The theoretical model
is considered in next section, where we present the dynamical model
of the system in Subsec. \ref{subsec:The-system}, the characterization
of the quantum states under consideration in Subsec. \ref{subsec:Characterization-of-the}
and the numerical methods in Subsec. \ref{subsec:Computational-method}.
In Sec. \ref{sec:Numerical-results} we present the numerical results
and our analyzes. We conclude the paper in Sec. \ref{sec:Conclusion}.

\section{Linear Array of Waveguides}

 The system we consider is a one-dimensional finite array of monomodal waveguides where prescribed input states of the electromagnetic field can propagate. Classically, fields propagating in the array of waveguides are coupled through evanescent waves passing over their boundary-barriers; on the quantum level, one says that photons can coherently tunnel between neighboring waveguides so that the quantum state represents the overall-overlapping superposition of the modes of the waveguides~\cite{Lahini_PRL08}. The field can be injected into one or a few waveguides and disorder can be implemented in the array either by randomly adjusting the spacing among the parallel waveguides along the $x$-direction, the propagation being in the $z$-direction, or by randomly fixing the thicknesses of the waveguides. This kind of system has been constructed on an AlGaAs substrate~\cite{Christodoulides_NAT03,Eisenberg_PRL98} and direct identification and measurements of Anderson-localization of states have been performed.

\subsection{Theoretical model}\label{subsec:The-system}

{

The electromagnetic energy density in the array of waveguides, assuming that all media are linear and nonmagnetic, and that the relevant evanescent overlap occurs only between neighboring waveguides, is given by
\begin{eqnarray}
{\mathcal H} & = & \sum_{j} \left( \frac12 \epsilon_0 n^2 {\mathbf E}_j^2 + \frac{1}{2 \mu_0} {\mathbf B}_j^2   \right) \nonumber \\
& & +\, \sum_{(j,j')} \left( \frac12 \epsilon_0 n^2 {\mathbf E}_j \cdot {\mathbf E}_{j'}  + \frac{1}{2\mu_0} {\mathbf B}_j \cdot {\mathbf B}_{j'}  \right) ,
\end{eqnarray}
where $n$ is the refraction index and $(j,j')$ denote next-neighbor pairs. We consider monochromatic fields propagating along the $z$-direction with velocity $c/n$.

At the quantum level, the field mode in the $j$-th waveguide is written in terms of photon-annihilation and photon-creation operators, $a_{j}$ and $a_{j}^{\dag}$ respectively, and the evolution of the system is dictated by a Hamiltonian in the form
\cite{Lahini_PRL10}
\begin{equation}
H = \sum_{j} \beta_{j} a_{j}^{\dag} a_{j} +   \sum_{j} \left[ C_{j+1,j} a_{j+1}^{\dag} a_{j} + C_{j,j+1} a_{j}^{\dag} a_{j+1} \right] ,\label{H}
\end{equation}
where $\beta_{j}$ is the propagation constant associated with the $j$-th waveguide and $C_{j+1,j}$ and $C_{j,j+1}$ are coupling coefficients between nearest neighbor waveguides. Notice that, we are considering fields with lower enough intensities to make non-linear effects negligible; also, along this paper unless stated in contrary, we use $\hbar=1$ and $c=1$.
The creation and annihilation operators satisfy the commutation relations
\begin{equation} [a_{j},a_{l}] = 0; \quad[a_{j}^{\dag},a_{l}^{\dag}] = 0;\quad[a_{j},a_{l}^{\dag}] = \delta_{jl} \, ,\label{aadag}
\end{equation}
and we assume the existence of eigenstates $|\psi\rangle$, such
that ${n}_{j}|\psi\rangle = a_{j}^{\dag}a_{j}|\psi\rangle = n_{j}|\psi\rangle$, where $n_{j}$ is the photon number of the $j$-th waveguide.

We will analyse arrays where we can fix a constant coupling (tunneling rate) between neighbor waveguides, $C_{j+1,j} = C_{j,j+1} = C$, and introduce disorder by taking the coefficients $\beta_{j}$ as random variables with zero-mean Gaussian distributions; the model then becomes isomorphic to the one-dimensional quantum tight-binding model used by Anderson~\cite{Anderson_PR58}, with $\beta_{j}$ being the on-site energy, and the system should then present localization of states. This assertion is experimentally feasible since these coefficients are related with the guide geometry, which can be appropriately adjusted~\cite{Lahini_PRL09,Lahini_PRA11} to spatially  modulate the index of refraction $n(x)$. In this case, the Heisenberg equations can be written as \cite{Bromberg_PRL09,Lahini_PRA11,Lahini_PRL08,Lahini_PRL09,Gilead_PRL15}
\begin{equation}
i\frac{\partial a_{j}}{\partial z} = [a_{j},H] = \beta_{j}a_{j} + C(a_{j+1}+a_{j-1}) ,\label{eqa}
\end{equation}
where $z=ct/n$, i.e. measurements of intensity distribution at position $z$ give the time evolution along the array.

Now we search for a solution of the Heisenberg equations~(\ref{eqa}), depending on the initial input state. Since the Heisenberg equations are linear in the annihilation (or for the creation) operators, it can be solved by finding the Green's function in such way that
\begin{equation}
a_{j}(z) = \sum_{l }G_{jl}(z) a_{l}(0) ,  \label{G}
\end{equation}
where $a_{l}(0)$ correspond to the input state (at $z=0$) into the $l$-th waveguide. The Green's function correlates fields in the $j$-th and the $l$-th waveguides at all positions $z$. By inserting Eq.~(\ref{G}) into Eq.~(\ref{eqa}), one gets the following set of
first order differential equations for the Green's functions,
\begin{equation}
i\frac{\partial G_{jl}}{\partial z} = \beta_{j}G_{jl} + C(G_{j+1,l}+G_{j-1,l}) .
\end{equation}
From our assumptions, $G_{jl}(z)$ depend on the parameters $\beta_{j}$, which vary randomly, and C that remains fixed; this is usually referred to as diagonal disorder. These equations can be solved numerically with great precision; taking specific distributions of $\{\beta_{j}\}$ and a given value of $C$, solutions are obtained just depending on the initial input state.

We can work with a great simplification if we consider that the input field {$|\Psi_{\text{in}}\rangle$} is injected into only one waveguide~\cite{Thompson_PRA10}, which we label by {$j_0$}, that is, { $ a_{j}(0) |\Psi_\text{in}\rangle= 0$ for all $j\neq j_0$}. 
{Thus, c}onsidering the light injection only into the {$j_0$-th} waveguide, the mean  field intensity {output by waveguide $j$ as a function of the lattice length $z$ is given by} 
\begin{equation}
\langle I_{j}(z) \rangle = \langle a_{j}^{\dag}(z)a_{j}(z) \rangle = \langle |G_{j,j_0}(z)|^{2} \rangle \langle a_{j_0}^{\dag}a_{j_0} \rangle , \label{I}
\end{equation}
where the mean value of the Green's functions is a standard statistical mean over several realizations for different values of $\{\beta_{j}\}$, while the mean {$\langle a_{j_0}^{\dag}a_{j_0}\rangle$} represents the quantum expectation value of the number operator {$a_{j_0}^{\dagger}{(0)} a_{j_0}{(0)}$},
which depends only on the input state. Also, we can calculate the correlation of the intensities between the output of two waveguides $j$ and $l$, {to be} given by
\begin{equation}
\langle I_{j}(z) I_{l}(z) \rangle = \langle |G_{j,{j_0}}(z)|^{2}|G_{l,{j_0}}(z)|^{2} \rangle \langle a_{{j_0}}^{\dag2} a_{{j_0}}^{2} \rangle.\label{II}
\end{equation}
Another important quantity to qualify the statistics of photons propagating in the array is the second-order correlation function
defined, for the $j$-th waveguide, by
\begin{equation}
g_{j}^{(2)}(z) = \frac{\langle a_{j}^{\dag2}(z) a_{j}^{2}(z) \rangle}{\langle a_{j}^{\dag}(z) a_{j}(z)\rangle^{2}} = \frac{\langle |G_{j,{j_0}}(z)|^{4} \rangle}{\langle |G_{j,{j_0}}(z)|^{2} \rangle^{2}}
\frac{\langle a_{{j_0}}^{\dag2} a_{{j_0}}^{2} \rangle}{\langle a_{{j_0}}^{\dag} a_{{j_0}} \rangle^{2}} .\label{g2}
\end{equation}
The correlation function $g^{(2)}(z)$ indicates weather states evolve  in a Poissonian way, or if they present either bunching or antibunching ($g^{(2)}(z) > 1$ and $g^{(2)}(z) < 1$, respectively) in their photon distribution. In any case, all these quantities depend on the initial state in the input of the array of waveguides. {The expressions for other quantities used to characterize the output field are presented in the results section.}
}

\subsection{Characterization of the {input} quantum states}\label{subsec:Characterization-of-the}

{
In Ref.~\cite{Thompson_PRA10}, a theoretical study of Anderson localization of light in an array of waveguides was presented, using coherent, thermal and squeezed states at the input, to investigate the effects of nonclassicality. Here, we extend the work of Ref.~\cite{Thompson_PRA10} by considering different input states that show more general quantum statistics, and by presenting the evolution in the Wigner representation. We shall consider specifically the complementary coherent state (${CCS}$)~\cite{Avelar_JOB04}, the
reciprocal binomial state (${RBS}$)~\cite{Valverde_PLA03}
and the polynomial state (${PS}$)~\cite{Souza_OC04} as input states; these states present peculiar statistical properties and it is interesting to investigate how they evolve along the array.

The complementary coherent state~\cite{Avelar_JOB04} is written in the number basis as
\begin{equation}
|\Psi_{CCS}(\alpha,N)\rangle = \aleph_{CCS}e^{-|\alpha|^{2}/2}\frac{2^{-N}}{\sqrt{N!}} \sum_{k=0}^{N}\sqrt{k!}\alpha^{*N-k}e^{ik\pi/2}|k\rangle ,\label{CCS}
\end{equation}
where the normalization constant $\aleph_{CCS}$ is given by
\begin{equation}
\aleph_{CCS}^{2}(\alpha,N) = \frac{N!e^{|\alpha|^{2}}2^{2N}}{\sum_{k=0}^{N}k!|\alpha|^{2(N-k)}} .
\end{equation}
For these states we find the mean number of photons,
\begin{equation}
\langle a_{{j_0}}^{\dag}a_{{j_0}}\rangle_{CCS} = \frac{\sum_{k=0}^{N}k!k|\alpha|^{2(N-k)}}{\sum_{k=0}^{N}k!|\alpha|^{2(N-k)}},\label{aadaga1}
\end{equation}
and the mean $\langle a_{{j_0}}^{\dag2}a_{{j_0}}^{2}\rangle$ is given by
\begin{equation}
\langle a_{{j_0}}^{\dag2}a_{{j_0}}^{2}\rangle_{CCS} = \frac{\sum_{k=0}^{N}k!k(k-1)|\alpha|^{2(N-k)}}{\sum_{k=0}^{N}k!|\alpha|^{2(N-k)}} .\label{aa21}
\end{equation}

The reciprocal binomial state \cite{Valverde_PLA03}, written as
\begin{equation}
|\Psi_{RBS}(\phi,N)\rangle = \aleph_{RBS}\sum_{k=0}^{N}\left(\begin{array}{c}
N\\
k
\end{array}\right)^{\frac{1}{2}} \exp\left[ik\left(\phi-\frac{\pi}{2}\right)\right]|k\rangle,\label{RBS}
\end{equation}
has normalization constant $\aleph_{RBS}$ given by
\begin{equation}
\aleph_{RBS}^{2}=\left[\sum_{k=0}^{N}\left(\begin{array}{c}
N\\
k
\end{array}\right)\right]^{-1} .
\end{equation}
For the $RBS$, we find
\begin{equation}
\langle a_{{j_0}}^{\dag}a_{{j_0}}\rangle_{RBS} = \aleph_{RBS}^{2} \sum_{k=0}^{N}\left(\begin{array}{c}
N\\
k
\end{array}\right)k \label{aadaga3}
\end{equation}
and
\begin{equation}
\langle a_{{j_0}}^{\dag2}a_{{j_0}}^{2}\rangle_{RBS} = \aleph_{RBS}^{2}\sum_{k=0}^{N}\left(\begin{array}{c}
N\\
k
\end{array}\right)k(k-1).\label{aa23}
\end{equation}

The polynomial state~\cite{Souza_OC04} is defined by
\begin{equation}
|\Psi_{PS}(x,N)\rangle = \aleph_{PS}\sum_{k=0}^{N}\left(\begin{array}{c}
N\\
k
\end{array}\right)^{-\frac{1}{2}} \frac{H_{N-k}(x/\sqrt{2})e^{\frac{ik\pi}{2}}}{\sqrt{(2N-2k-1)!!}} |k\rangle,\label{PS}
\end{equation}
where $H_{N-k}(y)$ is a Hermite polynomial, with the normalization constant $\aleph_{PS}$ written as
\begin{equation}
\aleph_{PS}^{2} = \left[\sum_{k=0}^{N}\left(\begin{array}{c}
N\\
k
\end{array}\right)^{-1} \frac{H_{N-k}^{2}(x/\sqrt{2})}{(2N-2k-1)!!}\right]^{-1} .
\end{equation}
For the polynomial state we find
\begin{equation}
\langle a_{{j_0}}^{\dag}a_{{j_0}}\rangle_{PS} = \aleph_{PS}^{2}\sum_{k=0}^{N}\left(\begin{array}{c}
N\\
k
\end{array}\right)^{-1}\frac{H_{N-k}^{2}(x/\sqrt{2})}{(2N-2k-1)!!}k \label{aadaga2}
\end{equation}
and
\begin{equation}
\langle a_{{j_0}}^{\dag 2} a_{{j_0}}^{2} \rangle_{PS} = \aleph_{PS}^{2}\sum_{k=0}^{N}\left(\begin{array}{c}
N\\
k
\end{array}\right)^{-1}\frac{H_{N-k}^{2}(x/\sqrt{2})}{(2N-2k-1)!!}k(k-1) .\label{aa22}
\end{equation}

{
For sake of comparison, we also investigate {the input states used in Ref. \cite{Thompson_PRA10}}: thermal states ($TS$), mixed states with density matrix
\begin{equation}
\rho_{TS} = \frac{1}{1+\bar{n}}\sum_{n=0}^{\infty} \left( \frac{\bar{n}}{1+\bar{n}} \right)^n \left| n \right\rangle \left\langle n \right| , \label{TS}
\end{equation}
for which $\langle a_{{j_0}}^{\dag}a_{{j_0}}\rangle_{TS} = \bar{n}$ and $\langle a_{{j_0}}^{\dag 2} a_{{j_0}}^{2} \rangle_{TS} = 2 \bar{n}^2$; coherent states ($CS$),
\begin{equation}
\left| \alpha \right\rangle = e^{-|\alpha|^2/2} \sum_{n=0}^{\infty} \frac{\alpha^n}{\sqrt{n!}}
\left| n \right\rangle , \label{CS}
\end{equation}
for which $\langle a_{{j_0}}^{\dag}a_{{j_0}}\rangle_{CS} = |\alpha|^2$ and $\langle a_{{j_0}}^{\dag 2} a_{{j_0}}^{2} \rangle_{CS} = |\alpha|^4$; and squeezed-vacuum states,
\begin{equation}
\left| \psi_{SS}(|\zeta|e^{\phi}) \right\rangle = \frac{1}{\cosh|\zeta|} \sum_{n=0}^{\infty} (-e^{i \phi} \tanh|\zeta|)^n \frac{\sqrt{(2 n)!}}{2^n\, n!} \left| 2 n \right\rangle ,
\end{equation}
with squeezing parameter $\zeta = |\zeta| e^{i\phi}$, for which $\langle a_{{j_0}}^{\dag}a_{{j_0}}\rangle_{SS} = \sinh^2|\zeta|$ and $\langle a_{{j_0}}^{\dag 2} a_{{j_0}}^{2} \rangle_{CS} = \sinh^2|\zeta| (1 + 3 \sinh^2|\zeta|)$. 
}

}

\subsection{Computational method}\label{subsec:Computational-method}

{
Since we are assuming that the input light is injected only in the 
{$j_0$-th} waveguide, all relevant quantities depend only on the Green's functions $G_{j,{j_0}}$, which satisfy
\begin{equation}
i\frac{\partial G_{j,{j_0}}}{\partial z} = \beta_{j}G_{j,{j_0}} + C(G_{j+1,{j_0}}+G_{j-1,{j_0}}) . \label{Gj0}
\end{equation}
To get solutions numerically, these equations were discretized and solved by using the Crank-Nicholson method {\citep{Crank_MPCPS47}} with step size $\Delta z=0.001$. In the numerical analysis, we consider an array with 101 waveguides, which we label from 1 to 101 with {light injected only in the 
waveguide $j_0=51$}. Thus the initial condition is that $G_{51,51}(0) = 1$ while $G_{j,51}(0) = 0$ for all $j \neq 51$; additionally, since the waveguide array is finite containing $M$ waveguides (here 101), we add {the} boundary conditions $G_{0,l}(z) = G_{M+1,l}(z) = 0$ for all $l = 1,2,....,M$.

Following Ref.~\cite{Thompson_PRA10}, we assume that the random coefficients $\beta_{j}$ are independent of each other and follow a zero-mean Gaussian probability distribution of the form
\begin{equation}
P(\beta) = \frac{1}{\sqrt{2 \pi \Delta^2}} \exp{\left( \frac{-\beta^2}{2\Delta^2} \right)} ,
\end{equation}
with the variance $\Delta^2$ measuring the disorder in the waveguide array. For sake of simplicity but without loss of generality, we fix the interaction parameter $C=1$ and take $\Delta/C$ quantifying the arrangement disorder.

In order to get the $\beta_{j}$ coefficients, we have used the Box-Müller method~\cite{Press} that requires the
generation of two random numbers with uniform distribution, which
was done using a congruence method. For each set $\{ \beta_j \}$ of disorder parameters, Eqs.~\eqref{Gj0} are solved numerically (fixing $C$) to find the relevant Green's functions. The functions $G_{j,{j_0}}(z)$ depend on the disorder parameters $\beta_j$ and on the coupling parameter $C$, and completely describe the dynamical evolution of any state injected into the waveguide array, only through the $51^{th}$ waveguide. Naturally, to average over the sets of the random coefficients $\beta_j$ one has to consider a reasonable number of realizations of disorder, different sets $\{ \beta_j \}$; here we take 1000 realizations of disorder in each simulation.

In our study, we take all the input states with the same energy, that is the same mean number of photons, specifically $n_{in}(0) = \langle a_{51}^{\dag}(0) a_{51}(0) \rangle = 10$. To present the results obtained by our simulations, we shall choose five states from the families we have shown in Subsec.~\ref{subsec:Characterization-of-the}, namely: two ${CCS}$ states, named ${CCS}_{1}$ ($ = \left| \Psi_{CCS}(0.1414,10) \right\rangle$) and ${CCS}_{2}$ ($ = \left| \Psi_{CCS}(1.916,11) \right\rangle$); two ${PS}$ states, referred to as
${PS}_{1}$ ($ = \left| \Psi_{PS}(0.374,12) \right\rangle $) and ${PS}_{2}$ ($ = \left| \Psi_{PS}(0.9345,13) \right\rangle $); and, in the case of the ${RBS}$ state, $\left| \Psi_{RBS} (\phi,N) \right\rangle$, we set $N=20$ {and $\phi = 0$}. {For completeness and comparison, we also consider as input states a thermal state ($TS$) with $\bar{n} = 10$, a coherent state ($CS$) with $|\alpha|^2 = 10$ and a squeezed-vacuum state ($SS$) with $\sinh^2 |\zeta| = 10$.}

}

\section{Numerical results}\label{sec:Numerical-results}

\begin{figure}[t]
\begin{centering}
\includegraphics[width=1\columnwidth]{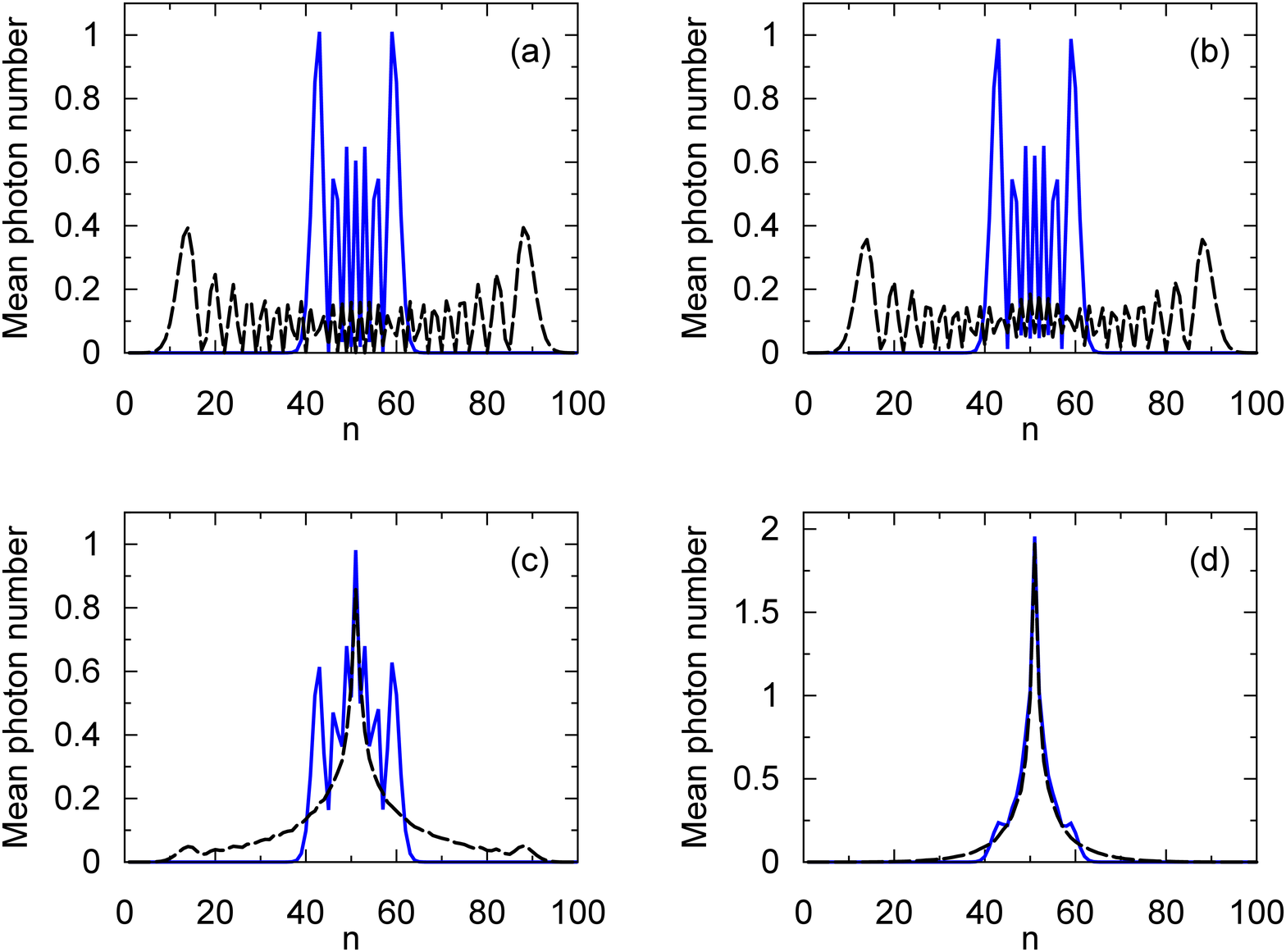}
\par\end{centering}
\caption{(Color online) {Mean photon number distribution at the output of the array of waveguides (numbered as $n = 1,2,3,\dots,101$) for two values of propagation distance
$z$: solid (blue) curves are for outputs at $z=5$, while
dashed (black) curves correspond to outputs at $z=20$. The disorder parameters were fixed as: (a)~$\Delta/C=0$, (b)~$\Delta/C=0.1$, (c)~$\Delta/C=0.5$, and (d)~$\Delta/C=1$. These plots were obtained by averaging over $1000$ realizations, fixing the number of photons in the injected (input) state ($z=0$, at the $51^{th}$ waveguide) $n_{in} = \langle a_{51}^{\dagger}(0) a_{51}(0) \rangle = 10$.}}

\label{F1}
\end{figure}

\begin{figure}[t]
\begin{centering}
\centering
\par\end{centering}
\begin{centering}
\includegraphics[width=1\columnwidth]{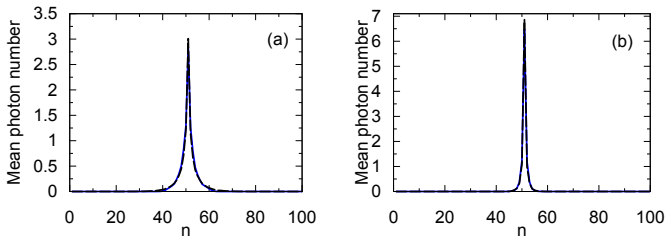}
\par\end{centering}
\caption{(Color online) The same as in Fig. \ref{F1}, but now using (a) $\Delta/C=1.5$
and (b) $\Delta/C=5$.}

\label{F2}
\end{figure}

{
As explicitly shown in Eq.~(\ref{I}), which is valid whenever light is injected {into} the 
 {$j_0$-th} waveguide, the average output intensity of the waveguide array depends on the average number of photons of the input state (here fixed as $n_{in}=10$) and on the degree of disorder of the system, carried by the random $\beta$-coefficients, which is manifested by the Green's functions; the output intensity, that is the mean number of photons at the end of the array, does not depend on any other characteristics of the input state.

The average output intensity, as distributed among the 101 waveguides of the array, is shown in Fig. \ref{F1} for some {values of} $\Delta/C$ and
two values of propagation distance $z$. As mentioned before, $\Delta/C$ measures the arrangement disorder, i.e., as the value of $\Delta$ increases the coefficients $\beta_{j}$ become more distinct from each other. In the absence of disorder, (Fig.~\ref{F1}(a)), the light suffers only the standard dispersion. However, as the disorder is increased (Fig.~\ref{F1}(b-d)), a narrow peak of intensity around the input ({$j_0$-th}, number 51) waveguide emerges, which characterizes the localization of the solution. For ratios $\Delta/C$ greater then $1.5$, the output-intensity profile does not change anymore with the increasing of the propagation distance, which is shown by the exact overlapping of the solid-line ($z=5$) and dashed-line ($z=20$) in Fig.~\ref{F2}.

}

\begin{figure}[t]
\begin{centering}
\centering \includegraphics[width=1\columnwidth]{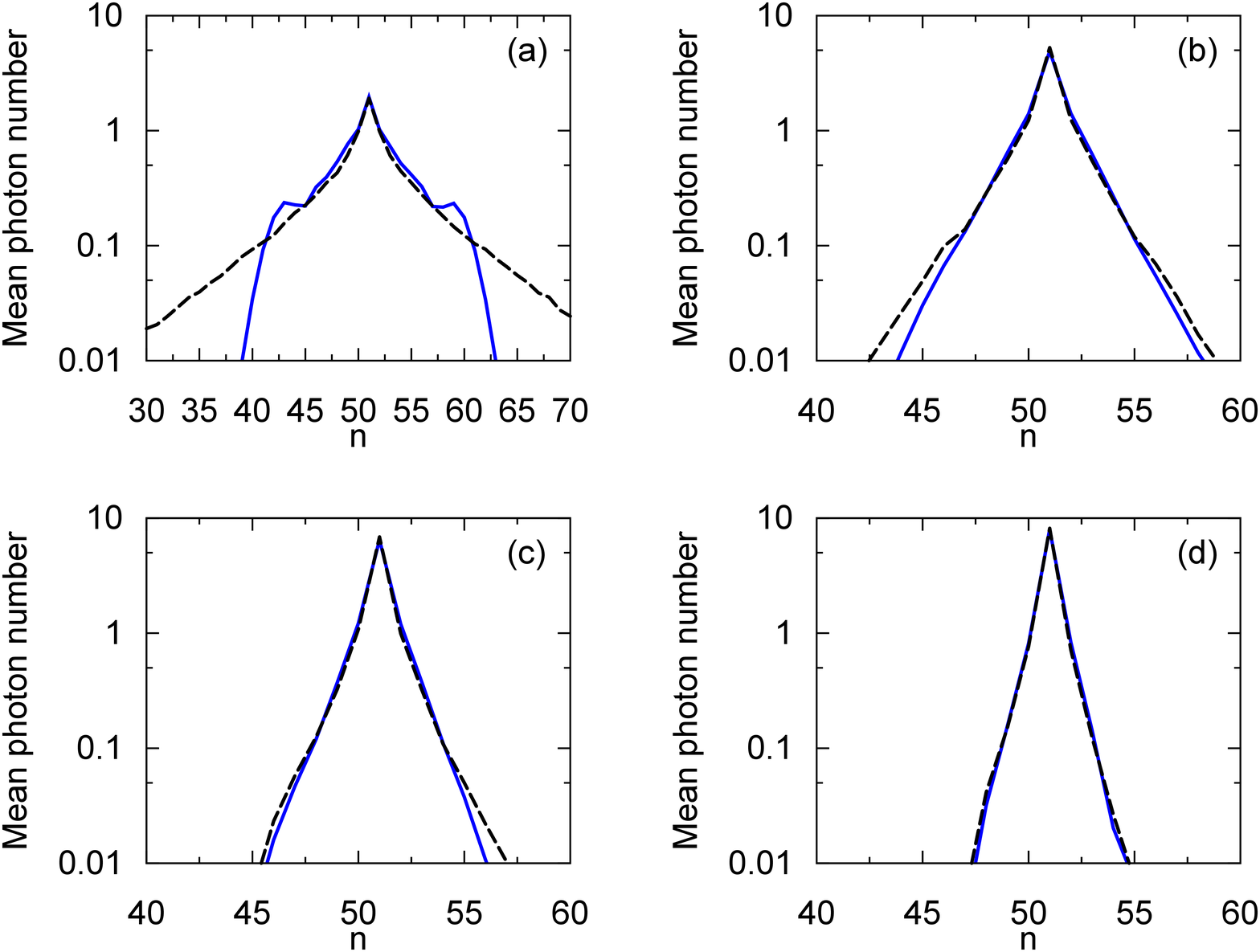}
\par\end{centering}
\caption{(Color online) The same as in Fig. \ref{F1}, but now in logarithm
scale and using the values (a) $\Delta/C=1$, (b) $\Delta/C=3$, (c)
$\Delta/C=5$, and (d) $\Delta/C=10$.}
\label{F3}
\end{figure}

{
In Fig.~\ref{F3} we present, using a logarithmic scale, the mean number of photons for some values of disorder ($\Delta/C$) and considering two values of propagation distance, $z=5$ and $z=20$ respectively. The choice of  using a logarithmic scale is {justified by the fact} that an exponential decay of the light intensity, in the waveguides different of the {$j_0$-th} one, would signalize the appearance of Anderson localization. We find that, in all cases, there exist two regions of decreasing exponentials and, except in case $\Delta/C=1.0$, the pattern holds for both propagation distances. It should be mentioned, however, that plots of the output intensities (mean number of photons) may not be sufficient to state whether the light beam presents Anderson localization, since the system might have yet a small diffusion, which would become more evident by increasing the propagation distance.

}

\begin{figure}[t]
\begin{centering}
\includegraphics[width=1\columnwidth]{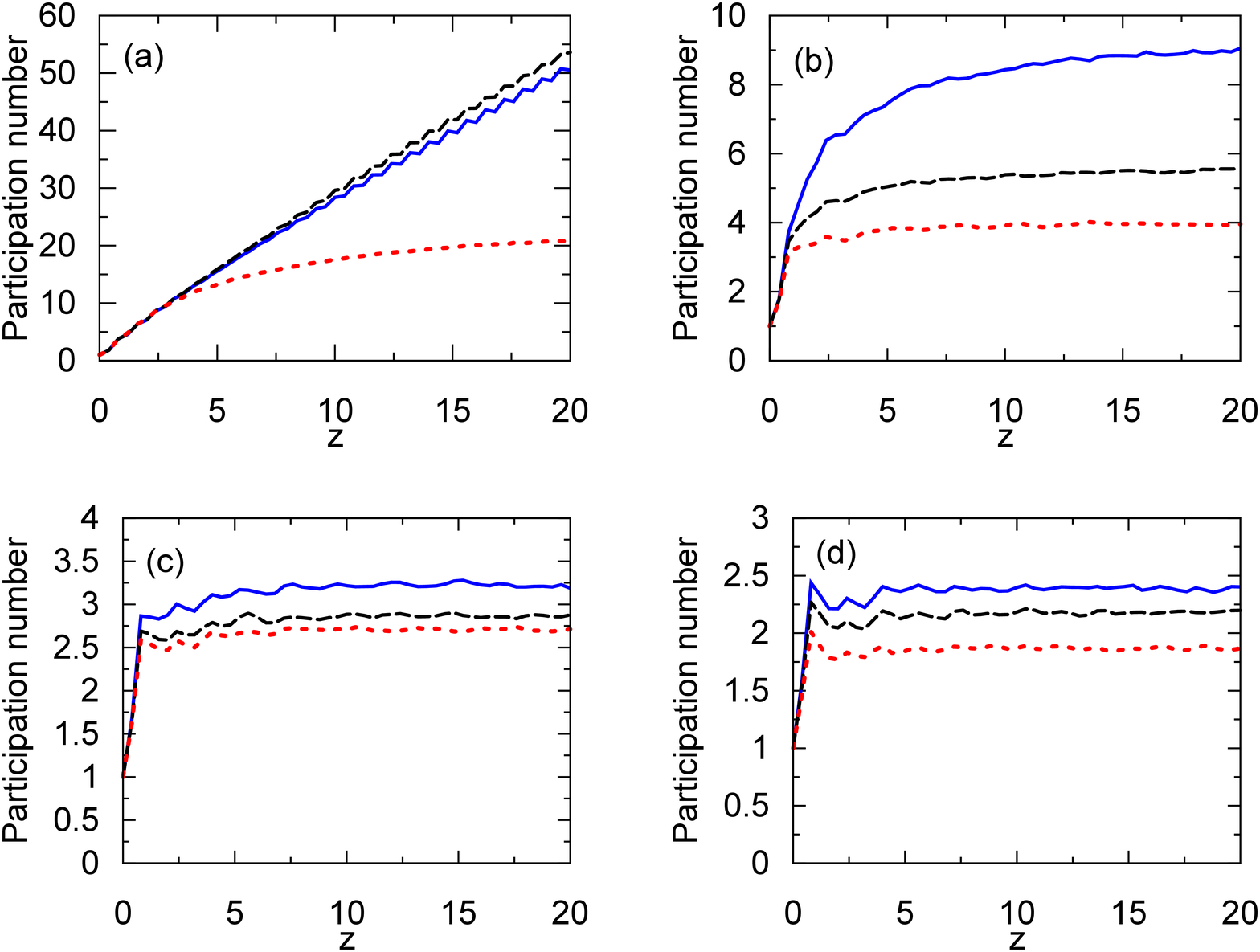}
\par\end{centering}
\caption{(Color online) {Participation number as a function of propagation distance
for various values of the disorder parameter: (a) $\Delta/C = 0$, $0.1$ and $0.5$; (b) $\Delta/C = 1$, $1.5$ and $2$; (c) $\Delta/C = 2.5$, $2.8$ and $3$; (d) $\Delta/C = 3.5$, $4$ and $5$; corresponding to solid-line (blue), dashed-line (black) and dotted-line (red), respectively, in all cases. Results were obtained after averaging over 1000 realizations.}}

\label{F4}
\end{figure}

{
In order to further investigate the occurrence of Anderson location, we can also calculate the participation number defined by, and given in our case by,
\begin{eqnarray}
\mathcal{P}(z) & = & \frac{\left(\sum_{j}\langle a_{j}^{\dag}(z) a_{j}(z) \rangle\right)^{2}}{\sum_{j}\langle a_{j}^{\dag}(z) a_{j}(z) \rangle^{2}} \nonumber \\
 & = & 1 + \frac{\sum_{j \neq k} \langle |G_{j,51}(z)|^{2} \rangle \langle |G_{k,51}(z)|^{2} \rangle}{\sum_{j} \langle |G_{j,51}(z)|^{2} \rangle^{2}} .
\label{Npar}
\end{eqnarray}
We see that the participation number is not only independent of the mean number of photons of the input state, but it is actually completely independent of the electromagnetic field mode injected in the {$j_0$-th} waveguide of the array, the number 51; it depends only on the waveguide array itself and its disorder encoded in the Green's functions.

The participation number indicates in how many waveguides there are
photons as a function of propagation distance; thus, if $\mathcal{P}(z)$ increases, it means that the beam remains scattering among the waveguides along propagation. It should be emphasized that, in the case of absence of disorder, the dispersion of $\mathcal{P}(z)$ is linear, as indicated by Fig.~\ref{F4}(a). Also, for a very small amount of disorder, $\Delta/C=0.1$, we find that the diffusion occurs faster than in the absence of disorder, while for $\Delta/C=0.5$ dispersion still occurs, but with a growth rate smaller than that in the case of no disorder. In all other cases reported in Figs.~\ref{F4}(b-d), for larger values of $\Delta/C$, the plots of $\mathcal{P}(z)$ stabilise as $z$ increases, characterizing localization of the light in the array. Notice that some fluctuations are still observed, but on a small scale when compared to $1$, which is the smallest value that can characterize a single waveguide.

}

\begin{figure}[t]
\begin{centering}
\includegraphics[width=1\columnwidth]{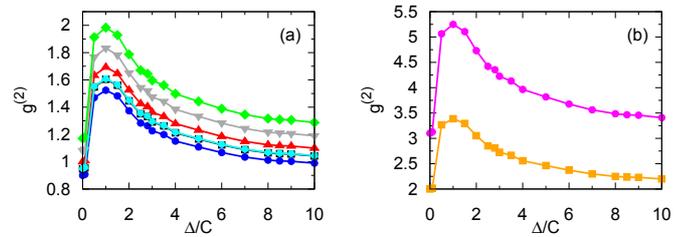}
\par\end{centering}
\caption{(Color online) {Second order correlation function of the output state in the $51^{th}$ waveguide, at propagation distance  $z=5$, as a function of the disorder parameter for different input states. Panel (a) displays the curves corresponding to the states ${CCS}_{1}$ in circles (blue), ${CCS}_{2}$ in boxes (black), ${RBS}$ in pentagon (cyan), $CS$ in triangles (red), ${PS}_{1}$ in down triangles (grey), and ${PS}_{2}$ in diamond (green). For comparison, panel (b) also presents the states $SS$ in circles (magenta) and $TS$ in boxes (orange). All results were obtained with an average over 1000 realizations.}}

\label{F5}
\end{figure}

\begin{figure}[t]
\includegraphics[width=1\columnwidth]{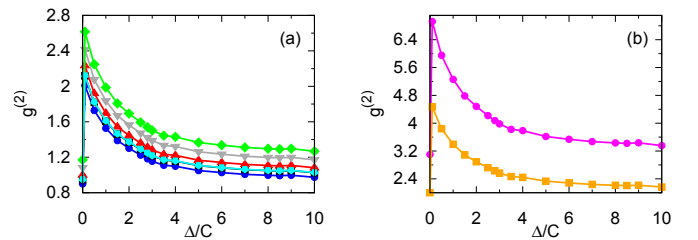}

\caption{(Color online) The same as in Fig.~\ref{F5}, but now with propagation distance $z=20$.}

\label{F6}
\end{figure}

{
Figures~\ref{F5} and \ref{F6} show the second order correlation function $g^{(2)}$ {of the field output by} waveguide number 51, as a function of the disorder degree for different input states, calculated with Eq.~(\ref{g2}), taking propagation distances $z=5$ and $z=20$, respectively.
We first notice that, for all input states analysed, taking measurements for a short propagation length ($z=5$, Fig.~\ref{F5}), the second order correlation function increases, as the disorder degree ($\Delta/C$) raises, it reaches a maximum for $\Delta/C=1$ and then decreases to values greater than the value of $g^{(2)}$ without disorder.
Similarly, for longer arrays ($z=20$, Fig.~\ref{F6}), one finds peaks of $g^{(2)}$ somewhat higher than those of Fig.~\ref{F5} but occurring at a much lower disorder degree and rapidly decaying for large disorder degree. Yet interestingly, for the propagation distance $z=20$, we find that only the ${CCS}_{1}$ state returns to the $g^{(2)}<1$ regime for large values of disorder. {Note also that the curves for the $CCS_2$ and the $RBS$ states are nearly coincident due to the fact that, for these states, $\langle a_{51}^{\dag 2}(0) a_{51}^{2}(0) \rangle$ are very close, $94.78$ and $95$, respectively.
}
}

\begin{figure}[tb]
\begin{centering}
\includegraphics[width=1\columnwidth]{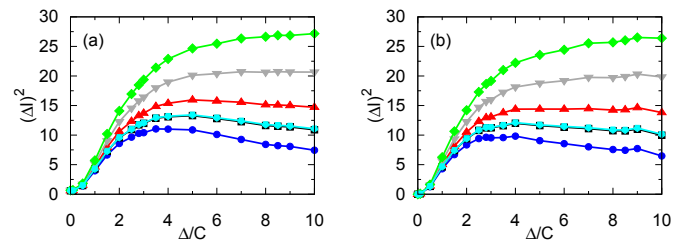}
\par\end{centering}

\caption{(Color online) {Variance of the output intensity of the $51^{th}$ waveguide as a function of the disorder parameter for the propagation distances
(a) $z=5$ and (b) $z=20$. We display the states ${CCS}_{1}$
in circles (blue), ${CCS}_{2}$ in boxes (black), ${RBS}$
in pentagon (cyan), $CS$ in triangles (red), ${PS}_{1}$ in down
triangles (grey), and ${PS}_{2}$ in diamond (green), respectively.}}

\label{F7}
\end{figure}

{
The variance of the output field intensity, $(\Delta I)^2 = \langle I^2 \rangle - \langle I \rangle^2$, of the $51^{th}$ waveguide, when the input field is injected only in it, is obtained directly from Eqs.~(\ref{I}) and \eqref{II} as
\begin{eqnarray}
(\Delta I_{51}(z))^{2} & = & \langle|G_{51,51}(z)|^{4}\rangle\langle a_{51}^{\dagger2}(0) a_{51}^{2}(0)\rangle \nonumber \\
  & & +\, \langle|G_{51,51}(z)|^{2}\rangle\langle a_{51}^{\dagger}(0) a_{51}(0) \rangle  \nonumber \\
 &  &  -\, \langle|G_{51,51}(z)|^{2}\rangle^{2}\langle a_{51}^{\dagger}(0) a_{51}(0)\rangle^{2} .\label{DI}
\end{eqnarray}

In order to investigate the influence of the disorder on the {intensity variance} of the output states, in Figs.~\ref{F7}(a) and \ref{F7}(b) we show these variances as functions of the disorder degree $\Delta/C$, at z = 5 and z = 20 respectively, for different input states injected into the $51^{th}$ waveguide.
We see that, for all states discussed, when the degree of disorder increases, the variances increase, some of them tending to stabilize, as for states ${PS}_{1}$, ${PS}_{2}$ and CS, while others, like the cases of the states ${CCS}_{1}$, ${CCS}_{2}$ and ${RBS}$, reach a maximum and then decrease for large values of the disorder parameter, the decreasing rate being bigger for the state ${CCS}_{1}$.
It is interesting to notice that the patterns of $(\Delta I)^{2}$ practically do not change when one compares the results for the propagation lengths z = 5 and z = 20, Figs.~\ref{F7}(a) and \ref{F7}(b) respectively. Here, likewise Figs.~\ref{F5} and \ref{F6}, the curves for the $CCS_2$ and the $RBS$ states are almost coincident. 

}

{
\subsection{Wigner representation}\label{WF}

We can also study the influence of disorder in the propagation of light in the waveguide array using the Wigner representation of the quantum states. The Wigner function (WF) of the output state in the $51^{th}$ waveguide is given by
\begin{equation}
W_{51}(\alpha,z) = \frac{1}{\pi} \int d^{2}\xi \exp(\alpha\xi^{*}-\alpha^{*}\xi)\, \chi_{51}(\xi,z) \, ,\label{wig}
\end{equation}
with the symmetrically ordered characteristic function given by
\begin{equation}
\chi_{51}(\xi,z) = {\mathrm Tr} \left[\rho_{51}(0)e^{\xi a_{51}^{\dagger}(z)-\xi^{*}a_{51}(z)}\right] , \label{chF}
\end{equation}
where $a_{51}(z) = G_{51,51}(z) a_{51}(0)$ and the input state is given by $\rho_{51}(0) = |\psi_{51}\rangle\langle\psi_{51}|$.

We take $|\psi_{51}\rangle = \sum_{n={0}}^{N} c_{n}|n\rangle$, which is the form of the states described in Subsec.~\ref{subsec:Characterization-of-the}.  Then, the integral defining $W_{51}(\alpha,z)$ can be performed analytically, and writing $\alpha=x+iy$, we obtain
\begin{widetext}
\begin{equation}
W_{51}^{(N)}(\alpha,z) = 2e^{-2(x^{2}+y^{2})} \sum_{t=0}^{N} \sum_{j=0}^{t} \sum_{n=0}^{N} \sum_{k=0}^{n} c_{n}^{*}c_{t} \left[ G_{51,51}(z) \right]^{j} \left[G_{51,51}^{*}(z)\right]^{k} \frac{\sqrt{t!n!}}{(n-k)!}\delta_{(n-k),(t-j)}I_{jk}(x,y),\label{Wigner}
\end{equation}
\end{widetext}
where
\begin{equation}
I_{jk}(x,y) = \sum_{l=0}^{k}\sum_{s=0}^{j} \frac{(-1)^{2j+k}\, i^{-(l+s)}}{(k-l)!l!(j-s)!s!}\, I_{p}(y) I_{q}(-x),
\end{equation}
with $p=l+s$ and $q=k-l+j-s$ and
\begin{equation}
I_{r}(u) = (2u)^{r} + \sum_{m=2}^{e(r)} \frac{(-1)^{\frac{m}{2}} 2^{-\frac{m}{2}} \,r!}{\left(\frac{m}{2}\right)!(r-m)!} \, (2u)^{r-m} ,
\end{equation}
where the summation is over even integers and $e(r)$ is the largest even integer not great{ter} than $r${, i.e. $e(r) = r$ if $r$ is even and $e(r) = r-1$, for $r$ odd.}
}

\begin{figure}[b]
\begin{centering}
\centering \includegraphics[width=0.48\columnwidth]{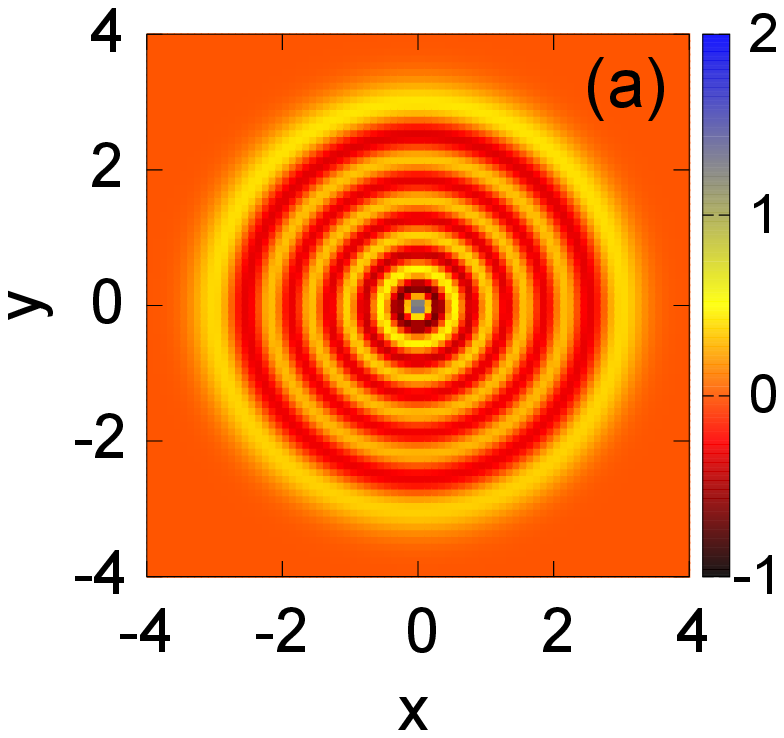} \includegraphics[width=0.48\columnwidth]{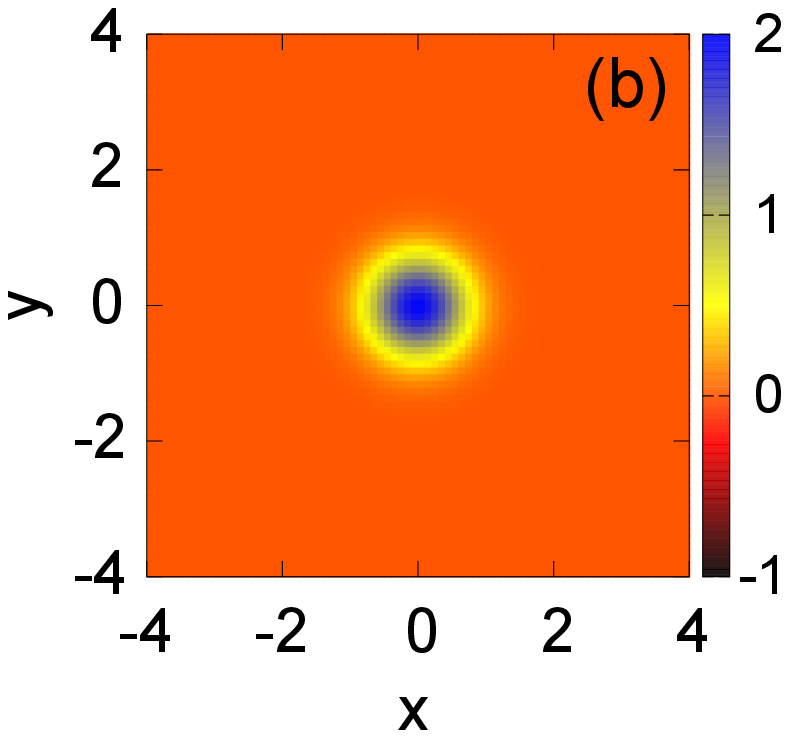}
\includegraphics[width=0.48\columnwidth]{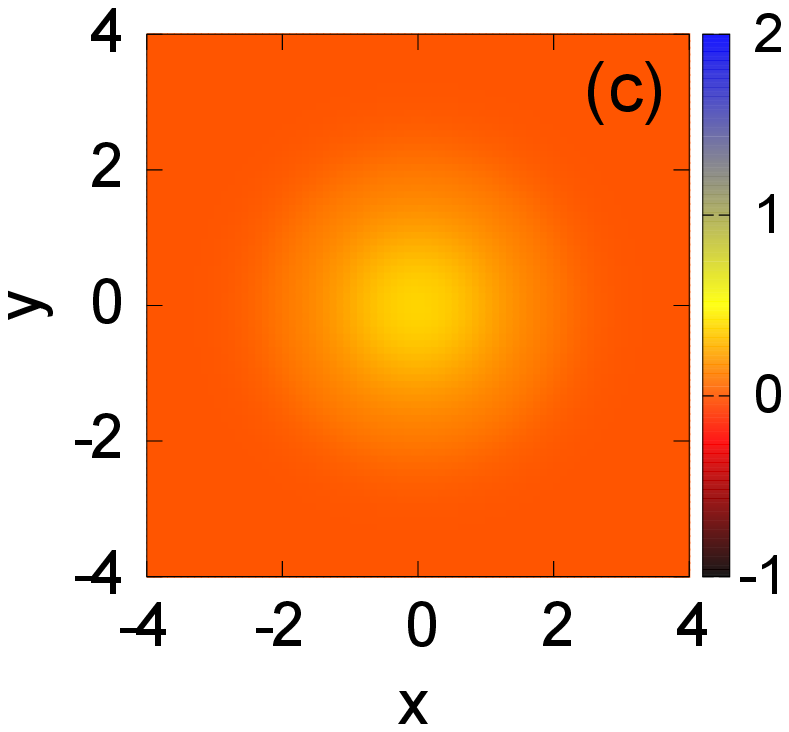} \includegraphics[width=0.48\columnwidth]{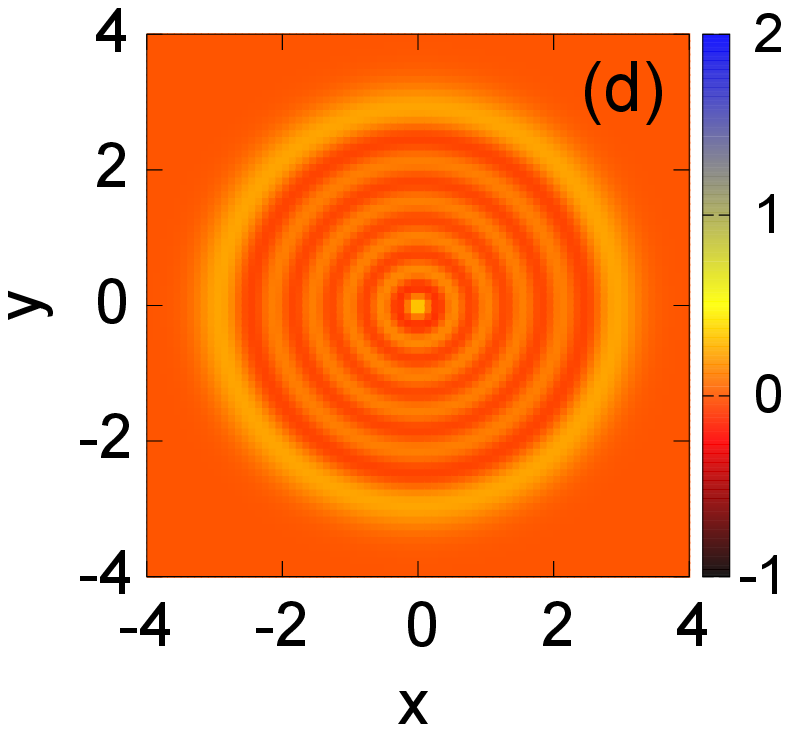}
\par\end{centering}
\caption{(Color online) {Panel (a) shows the Wigner function of the ${CCS}_{1}$ input state, while panels (b), (c) and (d) show the WF of the output state in the $51^{th}$ waveguide, at the propagation distance $z=20$, considering the disorder parameters $\Delta/C=0$, $1$ and $7$ respectively. The plots consist on an average over 100 realizations considering the step-size of $dx=dy=0.1$.}}

\label{F8}
\end{figure}

\begin{figure}[b]
\begin{centering}
\centering\includegraphics[width=0.48\columnwidth]{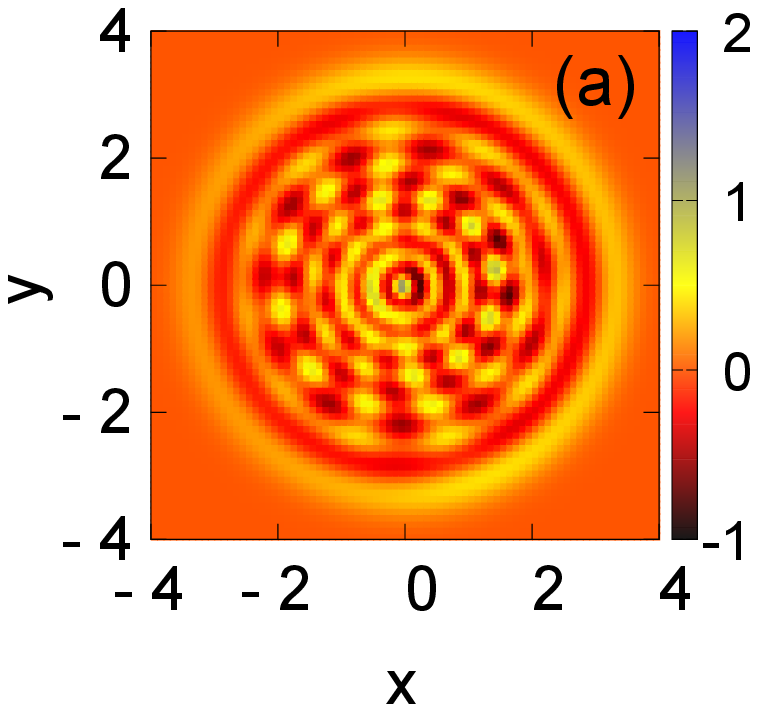} \includegraphics[width=0.48\columnwidth]{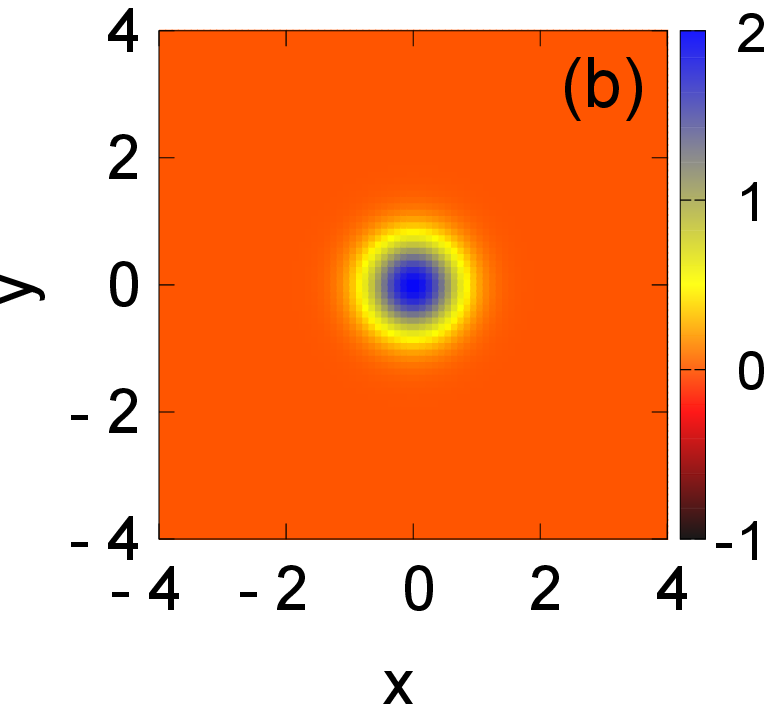}
\includegraphics[width=0.48\columnwidth]{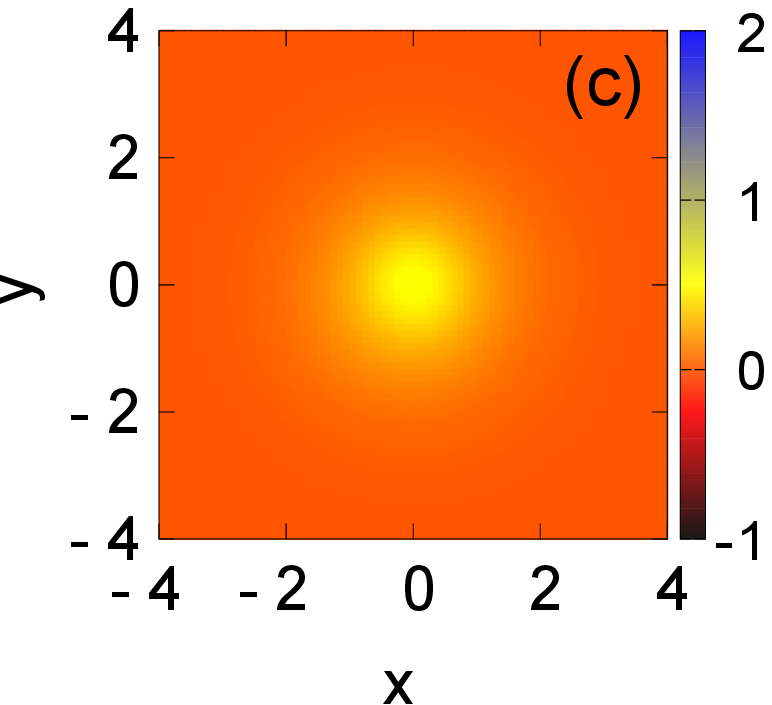} \includegraphics[width=0.48\columnwidth]{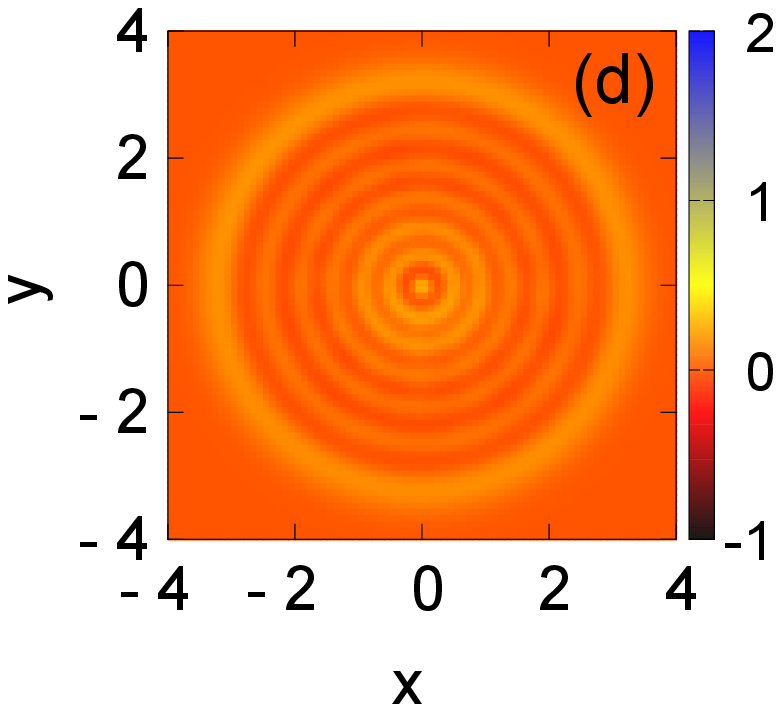}
\par\end{centering}
\caption{(Color online) {The same as in Fig. \ref{F8}, but now considering the ${PS}_{1}$ as the input state.}}

\label{F9}
\end{figure}

{
We can use the WF, given by Eq.~\eqref{Wigner} to investigate the propagation of states through the array, looking at the output state in the $51^{th}$ waveguide, when the input is a truncated state in the number basis, $\sum_{n=0}^{N} c_n |n\rangle$; Figs.~\ref{F8} and \ref{F9} show examples of this case. In Fig.~\ref{F8}, we present the WF when the input state is the ${CCS}_{1}$, in Fig.~\ref{F8}(a) for the input state ($z=0$), and in Figs.~\ref{F8}(b-d) for the output states in the $51^{th}$ waveguide, for the length $z=20$, considering the disorder parameters $\Delta/C=0$, $1$, and $7$, respectively.
We find that, in absence of disorder ($\Delta/C = 0$) (Fig.~\ref{F8}(b)), the WF is similar to the one of the vacuum state; this is also confirmed by the photon number distribution, as shown below. But, for small values of the disorder parameter, Figs.~\ref{F8}(b-c), the WF profiles present very important differences when compared with that for the input state{, showing that dispersion prevails}. However, by increasing the disorder parameter, the WF of the output state becomes similar to that one for the input state; clearly this behavior is directly linked to a strong localization, {but not necessarily of Anderson type}.
}

In Fig.~\ref{F9}(a), we present the WF for the input state ${PS}_{1}$
and, in Figs.~\ref{F9}(b-d), for the output state of the $51^{th}$
waveguide at $z=20$ with the same values of the disorder parameter
as in Fig.~\ref{F8}, $\Delta/C=0$, $1$, and $7$, respectively.
Here, we observed that, similarly to the results
obtained for the ${CCS}_{1}$ (Fig. \ref{F8}), when considering a
zero disorder (Fig.~\ref{F9}(b)) or a small disorder (here characterized
by the value of the disorder parameter $\Delta/C=1$, Fig.~\ref{F9}(c))
the output state of the $51^{th}$ waveguide at $z=20$ presents a
configuration for the WF similar to that of the vacuum state. Clearly,
the weak disorder is not enough to ensure that the WF remains in the
same shape, due to the interaction of the main guide ($51^{th}$)
with the neighboring guides. On the other hand, as the disorder parameter
increases (for example with $\Delta/C=7$ displayed in Fig.~\ref{F9}(d))
the WF is now preserved, which can be observed when comparing the
panels (a) and (d) of Figs. \ref{F8} and \ref{F9}. In other words,
by increasing the disorder parameter, the WFs become more robust to
changes due to variations of the coupling between neighbor waveguides.
Also, it should be pointed out that, in all cases, due to the computational
cost the number of realizations to produce the output state was reduced
to 100.

\begin{figure}[tb]
\begin{centering}
\centering \includegraphics[width=0.48\columnwidth]{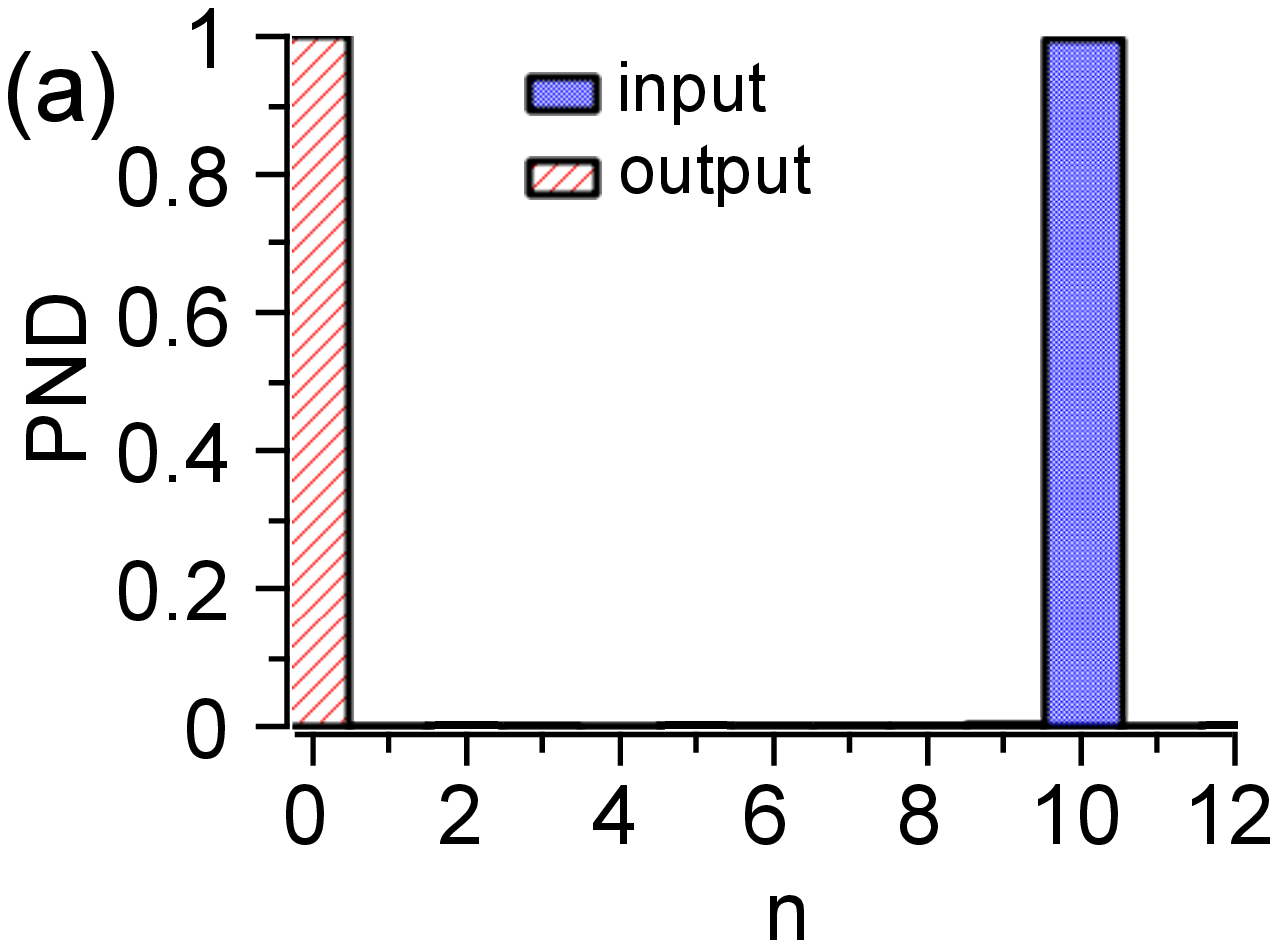} \includegraphics[width=0.48\columnwidth]{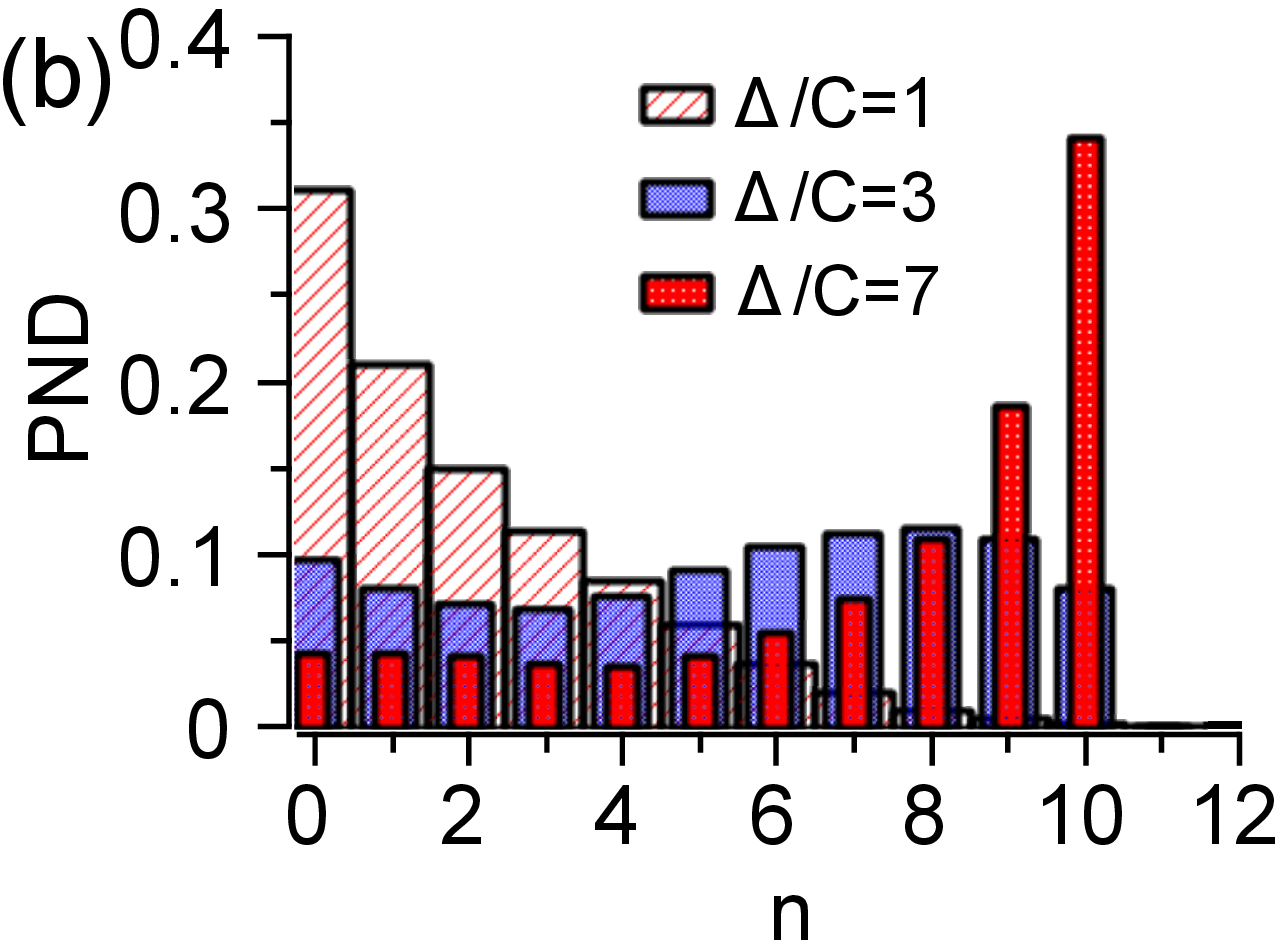}
\par\end{centering}
\caption{(Color online) {PND of the output state of the $51^{th}$ waveguide showing in (a) the input state ${CCS}_{1}$ and output state (at $z=20$) for $\Delta/C=0$ and, in (b), the output states for $\Delta/C=1$, $3$, and $7$. These results correspond to an average over $100$ realizations.}}

\label{F10}
\end{figure}

{
Finally, we can also use the WF (Eq.~(\ref{Wigner})) to obtain the
photon number distribution (PND) of the output state of the $51^{th}$ waveguide, given by
\begin{equation}
P_{51}^{(N)}(n,z) = \frac{1}{\pi} \int d^{2}\alpha\, W_{51}^{(N)}(\alpha,z)\, W_{n}(\alpha),\label{prob}
\end{equation}
where $W_{n}(\alpha)$ is the WF for the number state $|n\rangle$.
The integral in Eq.~(\ref{prob}) is calculated numerically and, as before, the number of realizations to get the proper averages was reduced to 100.

In Fig.~\ref{F10} we show the PND in the $51^{th}$ waveguide when the input state is the ${CCS}_{1}$ and {the length of the array} is $z=20$. In Fig.~\ref{F10}(a), we plot together the PND of the input ${CCS}_{1}$ state and the output PND ($P_{51}^{(N)}(n,20)$) for null disorder, for comparison. Interestingly, the PND of the state ${CCS}_{1}$, which has $n_{{CCS}_{1}}=10$, is very close to that of the number state $|10\rangle$, $P_{|10\rangle}(n) = \delta_{n,10}$. On the other hand, the output state of the $51^{th}$ waveguide (at $z=20$), in absence of disorder, is close to that for the vacuum state; this fact reinforces the results for the Wigner functions presented in Figs.~\ref{F8}(b) and \ref{F9}(b).
}

{
In Fig.~\ref{F10}(b) we show the PND for the output state in the $51^{th}$ waveguide considering three different values of the disorder parameter, $\Delta/C=0$, $1$, and $7$. We clearly see that the increasing of disorder tend to favor the output PND to become closer to that of the input state{, that is disorder tends to preserve the PND}.
}

\begin{figure}[tb]
\begin{centering}
\centering \includegraphics[width=0.48\columnwidth]{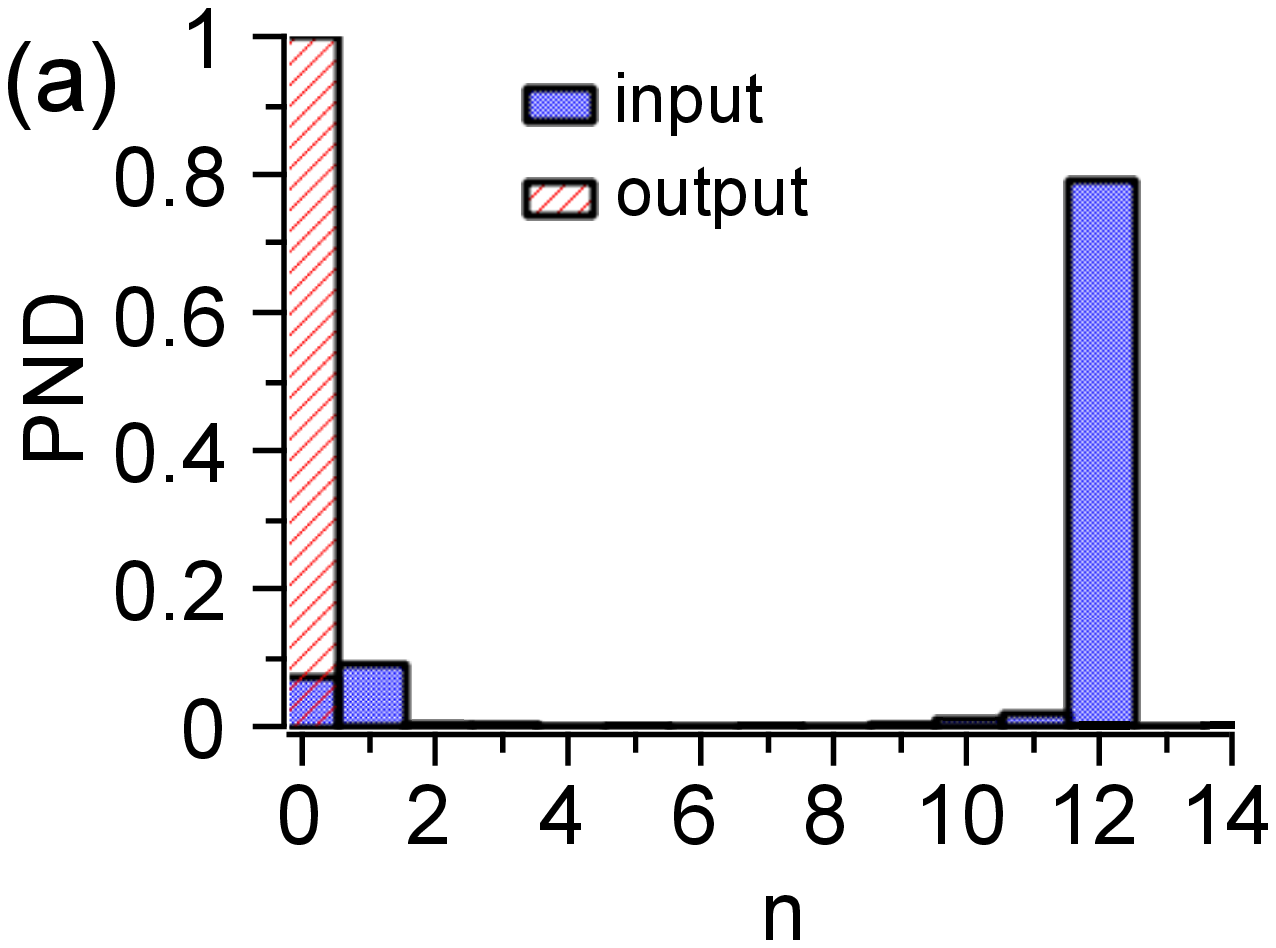} \includegraphics[width=0.48\columnwidth]{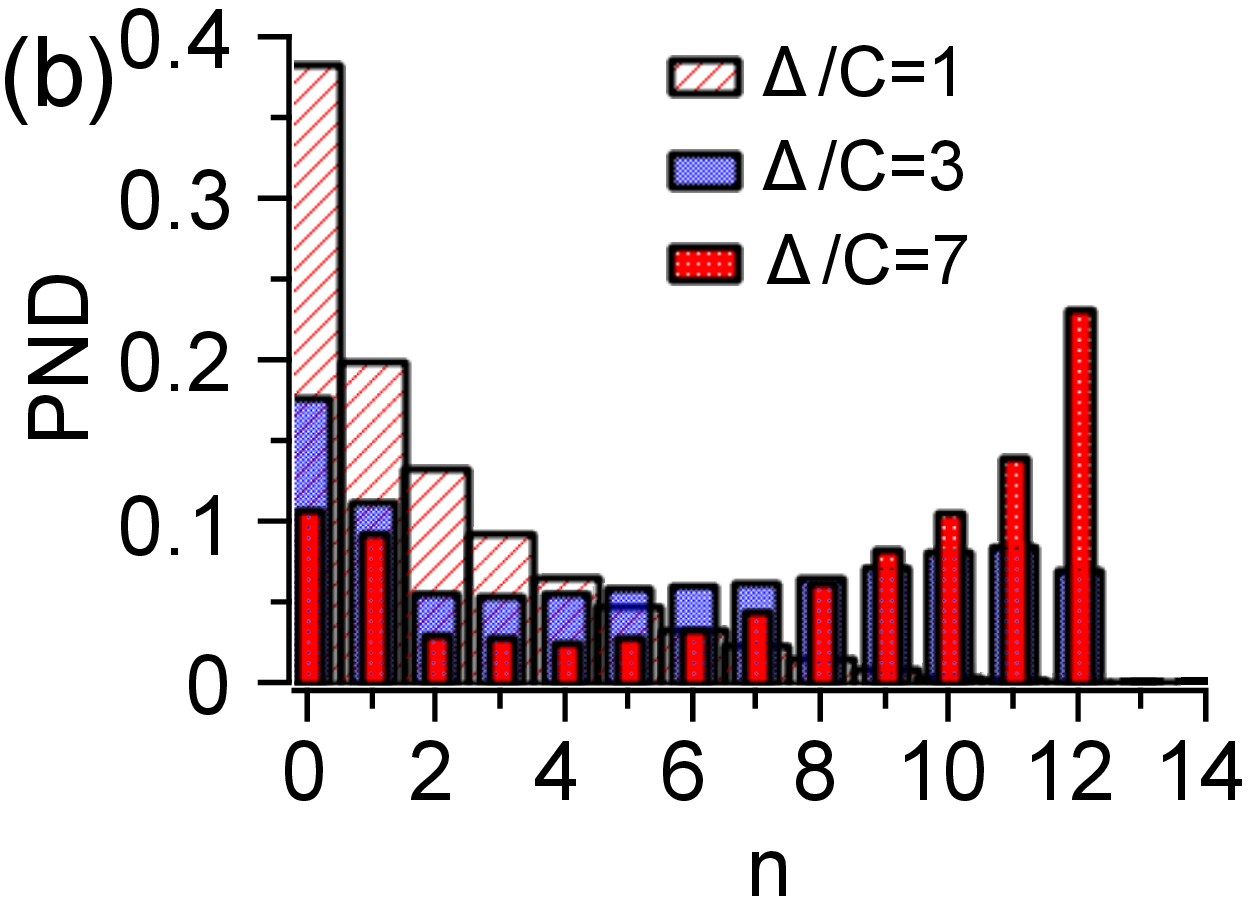}
\par\end{centering}
\caption{(Color online) The same as in Fig. \ref{F10} but now considering
the input state being the ${PS}_{1}$.}

\label{F11}
\end{figure}

In Fig. \ref{F11} we present the results of the PND, for the same parameters as in Fig.~\ref{F10}, but considering the input state given by the ${PS}_{1}$. Now we see, from Fig.~\ref{F11}(a), that the PND of the input ${PS}_{1}$ state {differs significantly} from the PND for the number state, distinctly to the case of the ${CCS}_{1}$, although we still have $n_{{PS}_{1}}=10$. However, also in agreement with the case of the ${CCS}_{1}$, we see in Fig. \ref{F11}(b) that the PND of the output state is clearly closer to that of the input state as greater is the value of $\Delta/C$.

\section{Conclusion}\label{sec:Conclusion}

We have discussed the evolution of quantum states of the electromagnetic field propagating through a disordered {plane} waveguide array. Specifically, we have analysed the propagation of three truncated states (in the number basis), namely the complementary coherent state, the reciprocal binomial state and the polynomial state; we also considered a thermal state, a coherent state and a squeezed-vacuum state, for comparison.

In our numerical calculations, we considered arrays with a 101 waveguides focusing in the injection and detection of waves in the middle one, the $51^{th}$-waveguide. The in site energy parameters $\beta_j$ were assumed to be independent of each other and randomly taken following a zero-mean Gaussian distribution with variance $\Delta^2$, while the coupling between neighbor waveguides were fixed as constant, $C$; the disorder parameter was defined as $\Delta/C$.

We have investigated some quantities that qualify the nature of the statistics of the state. First, we verified that increasing the degree of impurity of the lattice, the mean photon-number distribution tends to concentrate around the waveguide where the input state is injected, with an exponential decay of the light intensity in the others, for any injected state with a given mean number of photons, here $n_{in} = 10$; although this is not a definitive indication of localization, it does signalize it.

We also analysed the participation number, indicating in how many  waveguides there were photons, which shows the effects of disorder in the array and is totaly independent of the input state. We find, as the disorder parameter is increased, from no-disorder to a high disorder regime, the participation number changes from a linear increase with the propagation distance, characteristic of dispersion, to a flat behavior with very small fluctuations which represents a localization pattern.

{We investigated second order quantities as the $g^{(2)}$ function and the output intensity variance at the central waveguide. We observe for all input states that, although localized, the average output state presents a classical behavior relatively to the bunching feature. However, as the disorder increases, the $g^{(2)}$ function decreases and even presents antibunching again for the complementary coherent state and high disorder. While the variance in the output intensity at the central waveguide increases with disorder, and stabilizes to a final value for most of the input states investigated.

In order to investigate how the propagation through the lattice changes the input state, it is not enough to analyse quantities quantum averaged over the input state. It can be noticed, for example, when we look at the results for CCS2 and RBS, two different states with different features, however presenting nearly the same values of $g^{(2)}$ and intensity variance at the output.  To tackle this question we reconstructed the output state by means of its Wigner function, which could also be used to obtain the output probability distribution of number of photons. We observed a preservation of the characteristics of the Wigner function for high disorder parameter values, although we also notice a suppression of negative values in average. Moreover, as the output probability distribution of number of photons changes with the features of the array as well as with the input state, a well designed array of coupled waveguides could be used to produce new states of the electromagnetic field. }

\begin{acknowledgments}

We acknowledge financial support from the Brazilian agencies
CNPq (\#311408/2017-6, \#312723/2018-0, \#425718/2018-2 \& \#306065/2019-3), CAPES,
and FAPEG (PRONEM \#201710267000540, PRONEX \#201710267000503). This
work was performed as part of the Brazilian National Institute of
Science and Technology for Quantum Information (INCT-IQ \#465469/2014-0) and  Serrapilheira Institute.

\end{acknowledgments}

\bibliographystyle{apsrev4-1}
\bibliography{Refs}

\begin{thebibliography}{57}%
\makeatletter
\providecommand \@ifxundefined [1]{%
 \@ifx{#1\undefined}
}%
\providecommand \@ifnum [1]{%
 \ifnum #1\expandafter \@firstoftwo
 \else \expandafter \@secondoftwo
 \fi
}%
\providecommand \@ifx [1]{%
 \ifx #1\expandafter \@firstoftwo
 \else \expandafter \@secondoftwo
 \fi
}%
\providecommand \natexlab [1]{#1}%
\providecommand \enquote  [1]{``#1''}%
\providecommand \bibnamefont  [1]{#1}%
\providecommand \bibfnamefont [1]{#1}%
\providecommand \citenamefont [1]{#1}%
\providecommand \href@noop [0]{\@secondoftwo}%
\providecommand \href [0]{\begingroup \@sanitize@url \@href}%
\providecommand \@href[1]{\@@startlink{#1}\@@href}%
\providecommand \@@href[1]{\endgroup#1\@@endlink}%
\providecommand \@sanitize@url [0]{\catcode `\\12\catcode `\$12\catcode
  `\&12\catcode `\#12\catcode `\^12\catcode `\_12\catcode `\%12\relax}%
\providecommand \@@startlink[1]{}%
\providecommand \@@endlink[0]{}%
\providecommand \url  [0]{\begingroup\@sanitize@url \@url }%
\providecommand \@url [1]{\endgroup\@href {#1}{\urlprefix }}%
\providecommand \urlprefix  [0]{URL }%
\providecommand \Eprint [0]{\href }%
\providecommand \doibase [0]{http://dx.doi.org/}%
\providecommand \selectlanguage [0]{\@gobble}%
\providecommand \bibinfo  [0]{\@secondoftwo}%
\providecommand \bibfield  [0]{\@secondoftwo}%
\providecommand \translation [1]{[#1]}%
\providecommand \BibitemOpen [0]{}%
\providecommand \bibitemStop [0]{}%
\providecommand \bibitemNoStop [0]{.\EOS\space}%
\providecommand \EOS [0]{\spacefactor3000\relax}%
\providecommand \BibitemShut  [1]{\csname bibitem#1\endcsname}%
\let\auto@bib@innerbib\@empty
\bibitem [{\citenamefont {Bennett}\ \emph {et~al.}(1993)\citenamefont
  {Bennett}, \citenamefont {Brassard}, \citenamefont {Cr{\'{e}}peau},
  \citenamefont {Jozsa}, \citenamefont {Peres},\ and\ \citenamefont
  {Wootters}}]{Bennett_PRL93}%
  \BibitemOpen
  \bibfield  {author} {\bibinfo {author} {\bibfnamefont {C.~H.}\ \bibnamefont
  {Bennett}}, \bibinfo {author} {\bibfnamefont {G.}~\bibnamefont {Brassard}},
  \bibinfo {author} {\bibfnamefont {C.}~\bibnamefont {Cr{\'{e}}peau}}, \bibinfo
  {author} {\bibfnamefont {R.}~\bibnamefont {Jozsa}}, \bibinfo {author}
  {\bibfnamefont {A.}~\bibnamefont {Peres}}, \ and\ \bibinfo {author}
  {\bibfnamefont {W.~K.}\ \bibnamefont {Wootters}},\ }\href {\doibase
  10.1103/PhysRevLett.70.1895} {\bibfield  {journal} {\bibinfo  {journal}
  {Phys. Rev. Lett.}\ }\textbf {\bibinfo {volume} {70}},\ \bibinfo {pages}
  {1895} (\bibinfo {year} {1993})}\BibitemShut {NoStop}%
\bibitem [{\citenamefont {Bouwmeester}\ \emph {et~al.}(1997)\citenamefont
  {Bouwmeester}, \citenamefont {Pan}, \citenamefont {Mattle}, \citenamefont
  {Eibl}, \citenamefont {Weinfurter},\ and\ \citenamefont
  {Zeilinger}}]{Bouwmeester_NAT97}%
  \BibitemOpen
  \bibfield  {author} {\bibinfo {author} {\bibfnamefont {D.}~\bibnamefont
  {Bouwmeester}}, \bibinfo {author} {\bibfnamefont {J.-W.}\ \bibnamefont
  {Pan}}, \bibinfo {author} {\bibfnamefont {K.}~\bibnamefont {Mattle}},
  \bibinfo {author} {\bibfnamefont {M.}~\bibnamefont {Eibl}}, \bibinfo {author}
  {\bibfnamefont {H.}~\bibnamefont {Weinfurter}}, \ and\ \bibinfo {author}
  {\bibfnamefont {A.}~\bibnamefont {Zeilinger}},\ }\href {\doibase
  10.1038/37539} {\bibfield  {journal} {\bibinfo  {journal} {Nature}\ }\textbf
  {\bibinfo {volume} {390}},\ \bibinfo {pages} {575} (\bibinfo {year}
  {1997})}\BibitemShut {NoStop}%
\bibitem [{\citenamefont {Boschi}\ \emph {et~al.}(1998)\citenamefont {Boschi},
  \citenamefont {Branca}, \citenamefont {{De Martini}}, \citenamefont {Hardy},\
  and\ \citenamefont {Popescu}}]{Boschi_PRL98}%
  \BibitemOpen
  \bibfield  {author} {\bibinfo {author} {\bibfnamefont {D.}~\bibnamefont
  {Boschi}}, \bibinfo {author} {\bibfnamefont {S.}~\bibnamefont {Branca}},
  \bibinfo {author} {\bibfnamefont {F.}~\bibnamefont {{De Martini}}}, \bibinfo
  {author} {\bibfnamefont {L.}~\bibnamefont {Hardy}}, \ and\ \bibinfo {author}
  {\bibfnamefont {S.}~\bibnamefont {Popescu}},\ }\href {\doibase
  10.1103/PhysRevLett.80.1121} {\bibfield  {journal} {\bibinfo  {journal}
  {Phys. Rev. Lett.}\ }\textbf {\bibinfo {volume} {80}},\ \bibinfo {pages}
  {1121} (\bibinfo {year} {1998})}\BibitemShut {NoStop}%
\bibitem [{\citenamefont {Bennett}\ \emph {et~al.}(1992)\citenamefont
  {Bennett}, \citenamefont {Bessette}, \citenamefont {Brassard}, \citenamefont
  {Salvail},\ and\ \citenamefont {Smolin}}]{Bennett_JC92}%
  \BibitemOpen
  \bibfield  {author} {\bibinfo {author} {\bibfnamefont {C.~H.}\ \bibnamefont
  {Bennett}}, \bibinfo {author} {\bibfnamefont {F.}~\bibnamefont {Bessette}},
  \bibinfo {author} {\bibfnamefont {G.}~\bibnamefont {Brassard}}, \bibinfo
  {author} {\bibfnamefont {L.}~\bibnamefont {Salvail}}, \ and\ \bibinfo
  {author} {\bibfnamefont {J.}~\bibnamefont {Smolin}},\ }\href {\doibase
  10.1007/BF00191318} {\bibfield  {journal} {\bibinfo  {journal} {J. Cryptol.}\
  }\textbf {\bibinfo {volume} {5}},\ \bibinfo {pages} {3} (\bibinfo {year}
  {1992})}\BibitemShut {NoStop}%
\bibitem [{\citenamefont {Nielsen}\ and\ \citenamefont
  {Chuang}(2010)}]{Nielsen_10}%
  \BibitemOpen
  \bibfield  {author} {\bibinfo {author} {\bibfnamefont {M.~A.}\ \bibnamefont
  {Nielsen}}\ and\ \bibinfo {author} {\bibfnamefont {I.~L.}\ \bibnamefont
  {Chuang}},\ }\href {https://books.google.com.br/books?id=-s4DEy7o-a0C} {\emph
  {\bibinfo {title} {{Quantum Computation and Quantum Information: 10th
  Anniversary Edition}}}}\ (\bibinfo  {publisher} {Cambridge University
  Press},\ \bibinfo {year} {2010})\BibitemShut {NoStop}%
\bibitem [{\citenamefont {Kimble}(2008)}]{Kimble_NAT08}%
  \BibitemOpen
  \bibfield  {author} {\bibinfo {author} {\bibfnamefont {H.~J.}\ \bibnamefont
  {Kimble}},\ }\href {\doibase 10.1038/nature07127} {\bibfield  {journal}
  {\bibinfo  {journal} {Nature}\ }\textbf {\bibinfo {volume} {453}},\ \bibinfo
  {pages} {1023} (\bibinfo {year} {2008})}\BibitemShut {NoStop}%
\bibitem [{\citenamefont {Joannopoulos}\ \emph {et~al.}(2011)\citenamefont
  {Joannopoulos}, \citenamefont {Johnson}, \citenamefont {Winn},\ and\
  \citenamefont {Meade}}]{Joannopoulos_11}%
  \BibitemOpen
  \bibfield  {author} {\bibinfo {author} {\bibfnamefont {J.~D.}\ \bibnamefont
  {Joannopoulos}}, \bibinfo {author} {\bibfnamefont {S.~G.}\ \bibnamefont
  {Johnson}}, \bibinfo {author} {\bibfnamefont {J.~N.}\ \bibnamefont {Winn}}, \
  and\ \bibinfo {author} {\bibfnamefont {R.~D.}\ \bibnamefont {Meade}},\ }\href
  {https://books.google.com.br/books?id=owhE36qiTP8C} {\emph {\bibinfo {title}
  {{Photonic Crystals: Molding the Flow of Light - Second Edition}}}}\
  (\bibinfo  {publisher} {Princeton University Press},\ \bibinfo {year}
  {2011})\BibitemShut {NoStop}%
\bibitem [{\citenamefont {Joannopoulos}\ \emph {et~al.}(1997)\citenamefont
  {Joannopoulos}, \citenamefont {Villeneuve},\ and\ \citenamefont
  {Fan}}]{Joannopoulos_NAT97}%
  \BibitemOpen
  \bibfield  {author} {\bibinfo {author} {\bibfnamefont {J.~D.}\ \bibnamefont
  {Joannopoulos}}, \bibinfo {author} {\bibfnamefont {P.~R.}\ \bibnamefont
  {Villeneuve}}, \ and\ \bibinfo {author} {\bibfnamefont {S.}~\bibnamefont
  {Fan}},\ }\href {\doibase 10.1038/386143a0} {\bibfield  {journal} {\bibinfo
  {journal} {Nature}\ }\textbf {\bibinfo {volume} {386}},\ \bibinfo {pages}
  {143} (\bibinfo {year} {1997})}\BibitemShut {NoStop}%
\bibitem [{\citenamefont {Inoue}\ and\ \citenamefont
  {Ohtaka}(2004)}]{Inoue_04}%
  \BibitemOpen
  \bibfield  {author} {\bibinfo {author} {\bibfnamefont {K.}~\bibnamefont
  {Inoue}}\ and\ \bibinfo {author} {\bibfnamefont {K.}~\bibnamefont {Ohtaka}},\
  }\href {https://books.google.com.br/books?id=GIa3HRgPYhAC} {\emph {\bibinfo
  {title} {{Photonic Crystals: Physics, Fabrication and Applications}}}},\
  Springer Series in Optical Sciences\ (\bibinfo  {publisher} {Springer Berlin
  Heidelberg},\ \bibinfo {year} {2004})\BibitemShut {NoStop}%
\bibitem [{\citenamefont {Johnson}\ and\ \citenamefont
  {Joannopoulos}(2001)}]{Johnson_01}%
  \BibitemOpen
  \bibfield  {author} {\bibinfo {author} {\bibfnamefont {S.~G.}\ \bibnamefont
  {Johnson}}\ and\ \bibinfo {author} {\bibfnamefont {J.~D.}\ \bibnamefont
  {Joannopoulos}},\ }\href {https://books.google.com.br/books?id=LOZAsek9y7oC}
  {\emph {\bibinfo {title} {{Photonic Crystals: The Road from Theory to
  Practice}}}}\ (\bibinfo  {publisher} {Springer US},\ \bibinfo {year}
  {2001})\BibitemShut {NoStop}%
\bibitem [{\citenamefont {Ishizaki}\ \emph {et~al.}(2013)\citenamefont
  {Ishizaki}, \citenamefont {Koumura}, \citenamefont {Suzuki}, \citenamefont
  {Gondaira},\ and\ \citenamefont {Noda}}]{Ishizaki_NP13}%
  \BibitemOpen
  \bibfield  {author} {\bibinfo {author} {\bibfnamefont {K.}~\bibnamefont
  {Ishizaki}}, \bibinfo {author} {\bibfnamefont {M.}~\bibnamefont {Koumura}},
  \bibinfo {author} {\bibfnamefont {K.}~\bibnamefont {Suzuki}}, \bibinfo
  {author} {\bibfnamefont {K.}~\bibnamefont {Gondaira}}, \ and\ \bibinfo
  {author} {\bibfnamefont {S.}~\bibnamefont {Noda}},\ }\href {\doibase
  10.1038/nphoton.2012.341} {\bibfield  {journal} {\bibinfo  {journal} {Nat.
  Photonics}\ }\textbf {\bibinfo {volume} {7}},\ \bibinfo {pages} {133}
  (\bibinfo {year} {2013})}\BibitemShut {NoStop}%
\bibitem [{\citenamefont {Rinne}\ \emph {et~al.}(2008)\citenamefont {Rinne},
  \citenamefont {Garc{\'{i}}a-Santamar{\'{i}}a},\ and\ \citenamefont
  {Braun}}]{Rinne_NP08}%
  \BibitemOpen
  \bibfield  {author} {\bibinfo {author} {\bibfnamefont {S.~A.}\ \bibnamefont
  {Rinne}}, \bibinfo {author} {\bibfnamefont {F.}~\bibnamefont
  {Garc{\'{i}}a-Santamar{\'{i}}a}}, \ and\ \bibinfo {author} {\bibfnamefont
  {P.~V.}\ \bibnamefont {Braun}},\ }\href {\doibase 10.1038/nphoton.2007.252}
  {\bibfield  {journal} {\bibinfo  {journal} {Nat. Photonics}\ }\textbf
  {\bibinfo {volume} {2}},\ \bibinfo {pages} {52} (\bibinfo {year}
  {2008})}\BibitemShut {NoStop}%
\bibitem [{\citenamefont {Braun}\ \emph {et~al.}(2006)\citenamefont {Braun},
  \citenamefont {Rinne},\ and\ \citenamefont
  {Garc{\'{i}}a-Santamar{\'{i}}a}}]{Braun_AM06}%
  \BibitemOpen
  \bibfield  {author} {\bibinfo {author} {\bibfnamefont {P.~V.}\ \bibnamefont
  {Braun}}, \bibinfo {author} {\bibfnamefont {S.~A.}\ \bibnamefont {Rinne}}, \
  and\ \bibinfo {author} {\bibfnamefont {F.}~\bibnamefont
  {Garc{\'{i}}a-Santamar{\'{i}}a}},\ }\href {\doibase 10.1002/adma.200600769}
  {\bibfield  {journal} {\bibinfo  {journal} {Adv. Mater.}\ }\textbf {\bibinfo
  {volume} {18}},\ \bibinfo {pages} {2665} (\bibinfo {year}
  {2006})}\BibitemShut {NoStop}%
\bibitem [{\citenamefont {Wiersma}\ \emph {et~al.}(1997)\citenamefont
  {Wiersma}, \citenamefont {Bartolini}, \citenamefont {Lagendijk},\ and\
  \citenamefont {Righini}}]{Wiersma_NAT97}%
  \BibitemOpen
  \bibfield  {author} {\bibinfo {author} {\bibfnamefont {D.~S.}\ \bibnamefont
  {Wiersma}}, \bibinfo {author} {\bibfnamefont {P.}~\bibnamefont {Bartolini}},
  \bibinfo {author} {\bibfnamefont {A.}~\bibnamefont {Lagendijk}}, \ and\
  \bibinfo {author} {\bibfnamefont {R.}~\bibnamefont {Righini}},\ }\href
  {\doibase 10.1038/37757} {\bibfield  {journal} {\bibinfo  {journal} {Nature}\
  }\textbf {\bibinfo {volume} {390}},\ \bibinfo {pages} {671} (\bibinfo {year}
  {1997})}\BibitemShut {NoStop}%
\bibitem [{\citenamefont {Schwartz}\ \emph {et~al.}(2007)\citenamefont
  {Schwartz}, \citenamefont {Bartal}, \citenamefont {Fishman},\ and\
  \citenamefont {Segev}}]{Schwartz_NAT07}%
  \BibitemOpen
  \bibfield  {author} {\bibinfo {author} {\bibfnamefont {T.}~\bibnamefont
  {Schwartz}}, \bibinfo {author} {\bibfnamefont {G.}~\bibnamefont {Bartal}},
  \bibinfo {author} {\bibfnamefont {S.}~\bibnamefont {Fishman}}, \ and\
  \bibinfo {author} {\bibfnamefont {M.}~\bibnamefont {Segev}},\ }\href
  {\doibase 10.1038/nature05623} {\bibfield  {journal} {\bibinfo  {journal}
  {Nature}\ }\textbf {\bibinfo {volume} {446}},\ \bibinfo {pages} {52}
  (\bibinfo {year} {2007})}\BibitemShut {NoStop}%
\bibitem [{\citenamefont {Lahini}\ \emph {et~al.}(2008)\citenamefont {Lahini},
  \citenamefont {Avidan}, \citenamefont {Pozzi}, \citenamefont {Sorel},
  \citenamefont {Morandotti}, \citenamefont {Christodoulides},\ and\
  \citenamefont {Silberberg}}]{Lahini_PRL08}%
  \BibitemOpen
  \bibfield  {author} {\bibinfo {author} {\bibfnamefont {Y.}~\bibnamefont
  {Lahini}}, \bibinfo {author} {\bibfnamefont {A.}~\bibnamefont {Avidan}},
  \bibinfo {author} {\bibfnamefont {F.}~\bibnamefont {Pozzi}}, \bibinfo
  {author} {\bibfnamefont {M.}~\bibnamefont {Sorel}}, \bibinfo {author}
  {\bibfnamefont {R.}~\bibnamefont {Morandotti}}, \bibinfo {author}
  {\bibfnamefont {D.~N.}\ \bibnamefont {Christodoulides}}, \ and\ \bibinfo
  {author} {\bibfnamefont {Y.}~\bibnamefont {Silberberg}},\ }\href {\doibase
  10.1103/PhysRevLett.100.013906} {\bibfield  {journal} {\bibinfo  {journal}
  {Phys. Rev. Lett.}\ }\textbf {\bibinfo {volume} {100}},\ \bibinfo {pages}
  {013906} (\bibinfo {year} {2008})}\BibitemShut {NoStop}%
\bibitem [{\citenamefont {Garc{\'{i}}a}\ \emph {et~al.}(2011)\citenamefont
  {Garc{\'{i}}a}, \citenamefont {Sapienza}, \citenamefont {Toninelli},
  \citenamefont {L{\'{o}}pez},\ and\ \citenamefont {Wiersma}}]{Garcia_PRA11}%
  \BibitemOpen
  \bibfield  {author} {\bibinfo {author} {\bibfnamefont {P.~D.}\ \bibnamefont
  {Garc{\'{i}}a}}, \bibinfo {author} {\bibfnamefont {R.}~\bibnamefont
  {Sapienza}}, \bibinfo {author} {\bibfnamefont {C.}~\bibnamefont {Toninelli}},
  \bibinfo {author} {\bibfnamefont {C.}~\bibnamefont {L{\'{o}}pez}}, \ and\
  \bibinfo {author} {\bibfnamefont {D.~S.}\ \bibnamefont {Wiersma}},\ }\href
  {\doibase 10.1103/PhysRevA.84.023813} {\bibfield  {journal} {\bibinfo
  {journal} {Phys. Rev. A}\ }\textbf {\bibinfo {volume} {84}},\ \bibinfo
  {pages} {023813} (\bibinfo {year} {2011})}\BibitemShut {NoStop}%
\bibitem [{\citenamefont {Sapienza}\ \emph {et~al.}(2010)\citenamefont
  {Sapienza}, \citenamefont {Thyrrestrup}, \citenamefont {Stobbe},
  \citenamefont {Garcia}, \citenamefont {Smolka},\ and\ \citenamefont
  {Lodahl}}]{Sapienza_SCI10}%
  \BibitemOpen
  \bibfield  {author} {\bibinfo {author} {\bibfnamefont {L.}~\bibnamefont
  {Sapienza}}, \bibinfo {author} {\bibfnamefont {H.}~\bibnamefont
  {Thyrrestrup}}, \bibinfo {author} {\bibfnamefont {S.}~\bibnamefont {Stobbe}},
  \bibinfo {author} {\bibfnamefont {P.~D.}\ \bibnamefont {Garcia}}, \bibinfo
  {author} {\bibfnamefont {S.}~\bibnamefont {Smolka}}, \ and\ \bibinfo {author}
  {\bibfnamefont {P.}~\bibnamefont {Lodahl}},\ }\href {\doibase
  10.1126/science.1185080} {\bibfield  {journal} {\bibinfo  {journal} {Science
  (80-. ).}\ }\textbf {\bibinfo {volume} {327}},\ \bibinfo {pages} {1352}
  (\bibinfo {year} {2010})}\BibitemShut {NoStop}%
\bibitem [{\citenamefont {Liu}\ \emph {et~al.}(2014)\citenamefont {Liu},
  \citenamefont {Garcia}, \citenamefont {Ek}, \citenamefont {Gregersen},
  \citenamefont {Suhr}, \citenamefont {Schubert}, \citenamefont {M{\o}rk},
  \citenamefont {Stobbe},\ and\ \citenamefont {Lodahl}}]{Liu_NN14}%
  \BibitemOpen
  \bibfield  {author} {\bibinfo {author} {\bibfnamefont {J.}~\bibnamefont
  {Liu}}, \bibinfo {author} {\bibfnamefont {P.~D.}\ \bibnamefont {Garcia}},
  \bibinfo {author} {\bibfnamefont {S.}~\bibnamefont {Ek}}, \bibinfo {author}
  {\bibfnamefont {N.}~\bibnamefont {Gregersen}}, \bibinfo {author}
  {\bibfnamefont {T.}~\bibnamefont {Suhr}}, \bibinfo {author} {\bibfnamefont
  {M.}~\bibnamefont {Schubert}}, \bibinfo {author} {\bibfnamefont
  {J.}~\bibnamefont {M{\o}rk}}, \bibinfo {author} {\bibfnamefont
  {S.}~\bibnamefont {Stobbe}}, \ and\ \bibinfo {author} {\bibfnamefont
  {P.}~\bibnamefont {Lodahl}},\ }\href {\doibase 10.1038/nnano.2014.34}
  {\bibfield  {journal} {\bibinfo  {journal} {Nat. Nanotechnol.}\ }\textbf
  {\bibinfo {volume} {9}},\ \bibinfo {pages} {285} (\bibinfo {year}
  {2014})}\BibitemShut {NoStop}%
\bibitem [{\citenamefont {Rockstuhl}\ and\ \citenamefont
  {Scharf}(2013)}]{Rockstuhl_13}%
  \BibitemOpen
  \bibfield  {author} {\bibinfo {author} {\bibfnamefont {C.}~\bibnamefont
  {Rockstuhl}}\ and\ \bibinfo {author} {\bibfnamefont {T.}~\bibnamefont
  {Scharf}},\ }\href {\doibase 10.1007/978-3-642-32475-8} {\emph {\bibinfo
  {title} {{Amorphous Nanophotonics}}}},\ edited by\ \bibinfo {editor}
  {\bibfnamefont {C.}~\bibnamefont {Rockstuhl}}\ and\ \bibinfo {editor}
  {\bibfnamefont {T.}~\bibnamefont {Scharf}},\ Nano-Optics and Nanophotonics\
  (\bibinfo  {publisher} {Springer Berlin Heidelberg},\ \bibinfo {address}
  {Berlin, Heidelberg},\ \bibinfo {year} {2013})\BibitemShut {NoStop}%
\bibitem [{\citenamefont {Burgess}\ \emph {et~al.}(2016)\citenamefont
  {Burgess}, \citenamefont {Abedzadeh}, \citenamefont {Kay}, \citenamefont
  {Shneidman}, \citenamefont {Cranshaw}, \citenamefont {Lon{\v{c}}ar},\ and\
  \citenamefont {Aizenberg}}]{Burgess_SR16}%
  \BibitemOpen
  \bibfield  {author} {\bibinfo {author} {\bibfnamefont {I.~B.}\ \bibnamefont
  {Burgess}}, \bibinfo {author} {\bibfnamefont {N.}~\bibnamefont {Abedzadeh}},
  \bibinfo {author} {\bibfnamefont {T.~M.}\ \bibnamefont {Kay}}, \bibinfo
  {author} {\bibfnamefont {A.~V.}\ \bibnamefont {Shneidman}}, \bibinfo {author}
  {\bibfnamefont {D.~J.}\ \bibnamefont {Cranshaw}}, \bibinfo {author}
  {\bibfnamefont {M.}~\bibnamefont {Lon{\v{c}}ar}}, \ and\ \bibinfo {author}
  {\bibfnamefont {J.}~\bibnamefont {Aizenberg}},\ }\href {\doibase
  10.1038/srep19542} {\bibfield  {journal} {\bibinfo  {journal} {Sci. Rep.}\
  }\textbf {\bibinfo {volume} {6}},\ \bibinfo {pages} {19542} (\bibinfo {year}
  {2016})}\BibitemShut {NoStop}%
\bibitem [{\citenamefont {Anderson}(1958)}]{Anderson_PR58}%
  \BibitemOpen
  \bibfield  {author} {\bibinfo {author} {\bibfnamefont {P.~W.}\ \bibnamefont
  {Anderson}},\ }\href {\doibase 10.1103/PhysRev.109.1492} {\bibfield
  {journal} {\bibinfo  {journal} {Phys. Rev.}\ }\textbf {\bibinfo {volume}
  {109}},\ \bibinfo {pages} {1492} (\bibinfo {year} {1958})}\BibitemShut
  {NoStop}%
\bibitem [{\citenamefont {Chabanov}\ \emph {et~al.}(2000)\citenamefont
  {Chabanov}, \citenamefont {Stoytchev},\ and\ \citenamefont
  {Genack}}]{Chabanov_NAT00}%
  \BibitemOpen
  \bibfield  {author} {\bibinfo {author} {\bibfnamefont {A.~A.}\ \bibnamefont
  {Chabanov}}, \bibinfo {author} {\bibfnamefont {M.}~\bibnamefont {Stoytchev}},
  \ and\ \bibinfo {author} {\bibfnamefont {A.~Z.}\ \bibnamefont {Genack}},\
  }\href {\doibase 10.1038/35009055} {\bibfield  {journal} {\bibinfo  {journal}
  {Nature}\ }\textbf {\bibinfo {volume} {404}},\ \bibinfo {pages} {850}
  (\bibinfo {year} {2000})}\BibitemShut {NoStop}%
\bibitem [{\citenamefont {St{\"{o}}rzer}\ \emph {et~al.}(2006)\citenamefont
  {St{\"{o}}rzer}, \citenamefont {Gross}, \citenamefont {Aegerter},\ and\
  \citenamefont {Maret}}]{Storzer_PRL06}%
  \BibitemOpen
  \bibfield  {author} {\bibinfo {author} {\bibfnamefont {M.}~\bibnamefont
  {St{\"{o}}rzer}}, \bibinfo {author} {\bibfnamefont {P.}~\bibnamefont
  {Gross}}, \bibinfo {author} {\bibfnamefont {C.~M.}\ \bibnamefont {Aegerter}},
  \ and\ \bibinfo {author} {\bibfnamefont {G.}~\bibnamefont {Maret}},\ }\href
  {\doibase 10.1103/PhysRevLett.96.063904} {\bibfield  {journal} {\bibinfo
  {journal} {Phys. Rev. Lett.}\ }\textbf {\bibinfo {volume} {96}},\ \bibinfo
  {pages} {063904} (\bibinfo {year} {2006})}\BibitemShut {NoStop}%
\bibitem [{\citenamefont {Billy}\ \emph {et~al.}(2008)\citenamefont {Billy},
  \citenamefont {Josse}, \citenamefont {Zuo}, \citenamefont {Bernard},
  \citenamefont {Hambrecht}, \citenamefont {Lugan}, \citenamefont
  {Cl{\'{e}}ment}, \citenamefont {Sanchez-Palencia}, \citenamefont {Bouyer},\
  and\ \citenamefont {Aspect}}]{Billy_NAT08}%
  \BibitemOpen
  \bibfield  {author} {\bibinfo {author} {\bibfnamefont {J.}~\bibnamefont
  {Billy}}, \bibinfo {author} {\bibfnamefont {V.}~\bibnamefont {Josse}},
  \bibinfo {author} {\bibfnamefont {Z.}~\bibnamefont {Zuo}}, \bibinfo {author}
  {\bibfnamefont {A.}~\bibnamefont {Bernard}}, \bibinfo {author} {\bibfnamefont
  {B.}~\bibnamefont {Hambrecht}}, \bibinfo {author} {\bibfnamefont
  {P.}~\bibnamefont {Lugan}}, \bibinfo {author} {\bibfnamefont
  {D.}~\bibnamefont {Cl{\'{e}}ment}}, \bibinfo {author} {\bibfnamefont
  {L.}~\bibnamefont {Sanchez-Palencia}}, \bibinfo {author} {\bibfnamefont
  {P.}~\bibnamefont {Bouyer}}, \ and\ \bibinfo {author} {\bibfnamefont
  {A.}~\bibnamefont {Aspect}},\ }\href {\doibase 10.1038/nature07000}
  {\bibfield  {journal} {\bibinfo  {journal} {Nature}\ }\textbf {\bibinfo
  {volume} {453}},\ \bibinfo {pages} {891} (\bibinfo {year}
  {2008})}\BibitemShut {NoStop}%
\bibitem [{\citenamefont {Roati}\ \emph {et~al.}(2008)\citenamefont {Roati},
  \citenamefont {D'Errico}, \citenamefont {Fallani}, \citenamefont {Fattori},
  \citenamefont {Fort}, \citenamefont {Zaccanti}, \citenamefont {Modugno},
  \citenamefont {Modugno}, \citenamefont {Inguscio}, \citenamefont {D'Errico},
  \citenamefont {Fallani}, \citenamefont {Fattori}, \citenamefont {Fort},
  \citenamefont {Zaccanti}, \citenamefont {Modugno}, \citenamefont {Modugno},
  \citenamefont {Inguscio}, \citenamefont {D'Errico}, \citenamefont {Fallani},
  \citenamefont {Fattori}, \citenamefont {Fort}, \citenamefont {Zaccanti},
  \citenamefont {Modugno}, \citenamefont {Modugno}, \citenamefont {Inguscio},
  \citenamefont {D'Errico}, \citenamefont {Fallani}, \citenamefont {Fattori},
  \citenamefont {Fort}, \citenamefont {Zaccanti}, \citenamefont {Modugno},
  \citenamefont {Modugno},\ and\ \citenamefont {Inguscio}}]{Roati_NAT08}%
  \BibitemOpen
  \bibfield  {author} {\bibinfo {author} {\bibfnamefont {G.}~\bibnamefont
  {Roati}}, \bibinfo {author} {\bibfnamefont {C.}~\bibnamefont {D'Errico}},
  \bibinfo {author} {\bibfnamefont {L.}~\bibnamefont {Fallani}}, \bibinfo
  {author} {\bibfnamefont {M.}~\bibnamefont {Fattori}}, \bibinfo {author}
  {\bibfnamefont {C.}~\bibnamefont {Fort}}, \bibinfo {author} {\bibfnamefont
  {M.}~\bibnamefont {Zaccanti}}, \bibinfo {author} {\bibfnamefont
  {G.}~\bibnamefont {Modugno}}, \bibinfo {author} {\bibfnamefont
  {M.}~\bibnamefont {Modugno}}, \bibinfo {author} {\bibfnamefont
  {M.}~\bibnamefont {Inguscio}}, \bibinfo {author} {\bibfnamefont
  {C.}~\bibnamefont {D'Errico}}, \bibinfo {author} {\bibfnamefont
  {L.}~\bibnamefont {Fallani}}, \bibinfo {author} {\bibfnamefont
  {M.}~\bibnamefont {Fattori}}, \bibinfo {author} {\bibfnamefont
  {C.}~\bibnamefont {Fort}}, \bibinfo {author} {\bibfnamefont {M.}~\bibnamefont
  {Zaccanti}}, \bibinfo {author} {\bibfnamefont {G.}~\bibnamefont {Modugno}},
  \bibinfo {author} {\bibfnamefont {M.}~\bibnamefont {Modugno}}, \bibinfo
  {author} {\bibfnamefont {M.}~\bibnamefont {Inguscio}}, \bibinfo {author}
  {\bibfnamefont {C.}~\bibnamefont {D'Errico}}, \bibinfo {author}
  {\bibfnamefont {L.}~\bibnamefont {Fallani}}, \bibinfo {author} {\bibfnamefont
  {M.}~\bibnamefont {Fattori}}, \bibinfo {author} {\bibfnamefont
  {C.}~\bibnamefont {Fort}}, \bibinfo {author} {\bibfnamefont {M.}~\bibnamefont
  {Zaccanti}}, \bibinfo {author} {\bibfnamefont {G.}~\bibnamefont {Modugno}},
  \bibinfo {author} {\bibfnamefont {M.}~\bibnamefont {Modugno}}, \bibinfo
  {author} {\bibfnamefont {M.}~\bibnamefont {Inguscio}}, \bibinfo {author}
  {\bibfnamefont {C.}~\bibnamefont {D'Errico}}, \bibinfo {author}
  {\bibfnamefont {L.}~\bibnamefont {Fallani}}, \bibinfo {author} {\bibfnamefont
  {M.}~\bibnamefont {Fattori}}, \bibinfo {author} {\bibfnamefont
  {C.}~\bibnamefont {Fort}}, \bibinfo {author} {\bibfnamefont {M.}~\bibnamefont
  {Zaccanti}}, \bibinfo {author} {\bibfnamefont {G.}~\bibnamefont {Modugno}},
  \bibinfo {author} {\bibfnamefont {M.}~\bibnamefont {Modugno}}, \ and\
  \bibinfo {author} {\bibfnamefont {M.}~\bibnamefont {Inguscio}},\ }\href
  {\doibase 10.1038/nature07071} {\bibfield  {journal} {\bibinfo  {journal}
  {Nature}\ }\textbf {\bibinfo {volume} {453}},\ \bibinfo {pages} {895}
  (\bibinfo {year} {2008})}\BibitemShut {NoStop}%
\bibitem [{\citenamefont {Hu}\ \emph {et~al.}(2008)\citenamefont {Hu},
  \citenamefont {Strybulevych}, \citenamefont {Page}, \citenamefont
  {Skipetrov},\ and\ \citenamefont {van Tiggelen}}]{Hu_NP08}%
  \BibitemOpen
  \bibfield  {author} {\bibinfo {author} {\bibfnamefont {H.}~\bibnamefont
  {Hu}}, \bibinfo {author} {\bibfnamefont {A.}~\bibnamefont {Strybulevych}},
  \bibinfo {author} {\bibfnamefont {J.~H.}\ \bibnamefont {Page}}, \bibinfo
  {author} {\bibfnamefont {S.~E.}\ \bibnamefont {Skipetrov}}, \ and\ \bibinfo
  {author} {\bibfnamefont {B.~A.}\ \bibnamefont {van Tiggelen}},\ }\href
  {\doibase 10.1038/nphys1101} {\bibfield  {journal} {\bibinfo  {journal} {Nat.
  Phys.}\ }\textbf {\bibinfo {volume} {4}},\ \bibinfo {pages} {945} (\bibinfo
  {year} {2008})}\BibitemShut {NoStop}%
\bibitem [{\citenamefont {Srinivasan}\ \emph {et~al.}(2008)\citenamefont
  {Srinivasan}, \citenamefont {Aceves},\ and\ \citenamefont
  {Tartakovsky}}]{Srinivasan_PRA08}%
  \BibitemOpen
  \bibfield  {author} {\bibinfo {author} {\bibfnamefont {G.}~\bibnamefont
  {Srinivasan}}, \bibinfo {author} {\bibfnamefont {A.}~\bibnamefont {Aceves}},
  \ and\ \bibinfo {author} {\bibfnamefont {D.~M.}\ \bibnamefont
  {Tartakovsky}},\ }\href {\doibase 10.1103/PhysRevA.77.063806} {\bibfield
  {journal} {\bibinfo  {journal} {Phys. Rev. A}\ }\textbf {\bibinfo {volume}
  {77}},\ \bibinfo {pages} {063806} (\bibinfo {year} {2008})}\BibitemShut
  {NoStop}%
\bibitem [{\citenamefont {Cheng}\ and\ \citenamefont
  {Adhikari}(2010)}]{Cheng_PRA10}%
  \BibitemOpen
  \bibfield  {author} {\bibinfo {author} {\bibfnamefont {Y.}~\bibnamefont
  {Cheng}}\ and\ \bibinfo {author} {\bibfnamefont {S.~K.}\ \bibnamefont
  {Adhikari}},\ }\href {\doibase 10.1103/PhysRevA.82.013631} {\bibfield
  {journal} {\bibinfo  {journal} {Phys. Rev. A}\ }\textbf {\bibinfo {volume}
  {82}},\ \bibinfo {pages} {013631} (\bibinfo {year} {2010})}\BibitemShut
  {NoStop}%
\bibitem [{\citenamefont {Muruganandam}\ \emph {et~al.}(2010)\citenamefont
  {Muruganandam}, \citenamefont {Kumar},\ and\ \citenamefont
  {Adhikari}}]{Muruganandam_JPB10}%
  \BibitemOpen
  \bibfield  {author} {\bibinfo {author} {\bibfnamefont {P.}~\bibnamefont
  {Muruganandam}}, \bibinfo {author} {\bibfnamefont {R.~K.}\ \bibnamefont
  {Kumar}}, \ and\ \bibinfo {author} {\bibfnamefont {S.~K.}\ \bibnamefont
  {Adhikari}},\ }\href {\doibase 10.1088/0953-4075/43/20/205305} {\bibfield
  {journal} {\bibinfo  {journal} {J. Phys. B At. Mol. Opt. Phys.}\ }\textbf
  {\bibinfo {volume} {43}},\ \bibinfo {pages} {205305} (\bibinfo {year}
  {2010})}\BibitemShut {NoStop}%
\bibitem [{\citenamefont {Cheng}\ and\ \citenamefont
  {Adhikari}(2011{\natexlab{a}})}]{Cheng_PRA11}%
  \BibitemOpen
  \bibfield  {author} {\bibinfo {author} {\bibfnamefont {Y.}~\bibnamefont
  {Cheng}}\ and\ \bibinfo {author} {\bibfnamefont {S.~K.}\ \bibnamefont
  {Adhikari}},\ }\href {\doibase 10.1103/PhysRevA.83.023620} {\bibfield
  {journal} {\bibinfo  {journal} {Phys. Rev. A}\ }\textbf {\bibinfo {volume}
  {83}},\ \bibinfo {pages} {023620} (\bibinfo {year}
  {2011}{\natexlab{a}})}\BibitemShut {NoStop}%
\bibitem [{\citenamefont {Cheng}\ and\ \citenamefont
  {Adhikari}(2011{\natexlab{b}})}]{Cheng_PRA11-2}%
  \BibitemOpen
  \bibfield  {author} {\bibinfo {author} {\bibfnamefont {Y.}~\bibnamefont
  {Cheng}}\ and\ \bibinfo {author} {\bibfnamefont {S.~K.}\ \bibnamefont
  {Adhikari}},\ }\href {\doibase 10.1103/PhysRevA.84.023632} {\bibfield
  {journal} {\bibinfo  {journal} {Phys. Rev. A}\ }\textbf {\bibinfo {volume}
  {84}},\ \bibinfo {pages} {023632} (\bibinfo {year}
  {2011}{\natexlab{b}})}\BibitemShut {NoStop}%
\bibitem [{\citenamefont {Cheng}\ and\ \citenamefont
  {Adhikari}(2011{\natexlab{c}})}]{Cheng_PRA11-3}%
  \BibitemOpen
  \bibfield  {author} {\bibinfo {author} {\bibfnamefont {Y.}~\bibnamefont
  {Cheng}}\ and\ \bibinfo {author} {\bibfnamefont {S.~K.}\ \bibnamefont
  {Adhikari}},\ }\href {\doibase 10.1103/PhysRevA.84.053634} {\bibfield
  {journal} {\bibinfo  {journal} {Phys. Rev. A}\ }\textbf {\bibinfo {volume}
  {84}},\ \bibinfo {pages} {053634} (\bibinfo {year}
  {2011}{\natexlab{c}})}\BibitemShut {NoStop}%
\bibitem [{\citenamefont {Cardoso}\ \emph {et~al.}(2012)\citenamefont
  {Cardoso}, \citenamefont {Avelar},\ and\ \citenamefont
  {Bazeia}}]{Cardoso_NA12}%
  \BibitemOpen
  \bibfield  {author} {\bibinfo {author} {\bibfnamefont {W.~B.}\ \bibnamefont
  {Cardoso}}, \bibinfo {author} {\bibfnamefont {A.~T.}\ \bibnamefont {Avelar}},
  \ and\ \bibinfo {author} {\bibfnamefont {D.}~\bibnamefont {Bazeia}},\ }\href
  {\doibase 10.1016/j.nonrwa.2011.08.014} {\bibfield  {journal} {\bibinfo
  {journal} {Nonlinear Anal. Real World Appl.}\ }\textbf {\bibinfo {volume}
  {13}},\ \bibinfo {pages} {755} (\bibinfo {year} {2012})}\BibitemShut
  {NoStop}%
\bibitem [{\citenamefont {Cheng}\ \emph {et~al.}(2014)\citenamefont {Cheng},
  \citenamefont {Tang},\ and\ \citenamefont {Adhikari}}]{Cheng_PRA14}%
  \BibitemOpen
  \bibfield  {author} {\bibinfo {author} {\bibfnamefont {Y.}~\bibnamefont
  {Cheng}}, \bibinfo {author} {\bibfnamefont {G.}~\bibnamefont {Tang}}, \ and\
  \bibinfo {author} {\bibfnamefont {S.~K.}\ \bibnamefont {Adhikari}},\ }\href
  {\doibase 10.1103/PhysRevA.89.063602} {\bibfield  {journal} {\bibinfo
  {journal} {Phys. Rev. A}\ }\textbf {\bibinfo {volume} {89}},\ \bibinfo
  {pages} {063602} (\bibinfo {year} {2014})}\BibitemShut {NoStop}%
\bibitem [{\citenamefont {Xi}\ \emph {et~al.}(2015)\citenamefont {Xi},
  \citenamefont {Li},\ and\ \citenamefont {Shi}}]{Xi_PB15}%
  \BibitemOpen
  \bibfield  {author} {\bibinfo {author} {\bibfnamefont {K.-T.}\ \bibnamefont
  {Xi}}, \bibinfo {author} {\bibfnamefont {J.}~\bibnamefont {Li}}, \ and\
  \bibinfo {author} {\bibfnamefont {D.-N.}\ \bibnamefont {Shi}},\ }\href
  {\doibase 10.1016/j.physb.2014.11.068} {\bibfield  {journal} {\bibinfo
  {journal} {Phys. B Condens. Matter}\ }\textbf {\bibinfo {volume} {459}},\
  \bibinfo {pages} {6} (\bibinfo {year} {2015})}\BibitemShut {NoStop}%
\bibitem [{\citenamefont {Cardoso}\ \emph {et~al.}(2016)\citenamefont
  {Cardoso}, \citenamefont {Le{\~{a}}o},\ and\ \citenamefont
  {Avelar}}]{Cardoso_OQE16}%
  \BibitemOpen
  \bibfield  {author} {\bibinfo {author} {\bibfnamefont {W.~B.}\ \bibnamefont
  {Cardoso}}, \bibinfo {author} {\bibfnamefont {S.~A.}\ \bibnamefont
  {Le{\~{a}}o}}, \ and\ \bibinfo {author} {\bibfnamefont {A.~T.}\ \bibnamefont
  {Avelar}},\ }\href {\doibase 10.1007/s11082-016-0658-z} {\bibfield  {journal}
  {\bibinfo  {journal} {Opt. Quantum Electron.}\ }\textbf {\bibinfo {volume}
  {48}},\ \bibinfo {pages} {388} (\bibinfo {year} {2016})}\BibitemShut
  {NoStop}%
\bibitem [{\citenamefont {Cardoso}(2019)}]{Cardoso_PLA19}%
  \BibitemOpen
  \bibfield  {author} {\bibinfo {author} {\bibfnamefont {W.~B.}\ \bibnamefont
  {Cardoso}},\ }\href {\doibase 10.1016/j.physleta.2019.125898} {\bibfield
  {journal} {\bibinfo  {journal} {Phys. Lett. A}\ }\textbf {\bibinfo {volume}
  {383}},\ \bibinfo {pages} {125898} (\bibinfo {year} {2019})}\BibitemShut
  {NoStop}%
\bibitem [{\citenamefont {dos Santos}\ and\ \citenamefont
  {Cardoso}(2020)}]{Santos_ND20}%
  \BibitemOpen
  \bibfield  {author} {\bibinfo {author} {\bibfnamefont {M.~C.~P.}\
  \bibnamefont {dos Santos}}\ and\ \bibinfo {author} {\bibfnamefont {W.~B.}\
  \bibnamefont {Cardoso}},\ }\href {\doibase 10.1007/s11071-020-05788-z}
  {\bibfield  {journal} {\bibinfo  {journal} {Nonlinear Dyn.}\ }\textbf
  {\bibinfo {volume} {101}},\ \bibinfo {pages} {611} (\bibinfo {year}
  {2020})}\BibitemShut {NoStop}%
\bibitem [{\citenamefont {Thompson}\ \emph {et~al.}(2010)\citenamefont
  {Thompson}, \citenamefont {Vemuri},\ and\ \citenamefont
  {Agarwal}}]{Thompson_PRA10}%
  \BibitemOpen
  \bibfield  {author} {\bibinfo {author} {\bibfnamefont {C.}~\bibnamefont
  {Thompson}}, \bibinfo {author} {\bibfnamefont {G.}~\bibnamefont {Vemuri}}, \
  and\ \bibinfo {author} {\bibfnamefont {G.~S.}\ \bibnamefont {Agarwal}},\
  }\href {\doibase 10.1103/PhysRevA.82.053805} {\bibfield  {journal} {\bibinfo
  {journal} {Phys. Rev. A}\ }\textbf {\bibinfo {volume} {82}},\ \bibinfo
  {pages} {053805} (\bibinfo {year} {2010})}\BibitemShut {NoStop}%
\bibitem [{\citenamefont {Lahini}\ \emph {et~al.}(2009)\citenamefont {Lahini},
  \citenamefont {Pugatch}, \citenamefont {Pozzi}, \citenamefont {Sorel},
  \citenamefont {Morandotti}, \citenamefont {Davidson},\ and\ \citenamefont
  {Silberberg}}]{Lahini_PRL09}%
  \BibitemOpen
  \bibfield  {author} {\bibinfo {author} {\bibfnamefont {Y.}~\bibnamefont
  {Lahini}}, \bibinfo {author} {\bibfnamefont {R.}~\bibnamefont {Pugatch}},
  \bibinfo {author} {\bibfnamefont {F.}~\bibnamefont {Pozzi}}, \bibinfo
  {author} {\bibfnamefont {M.}~\bibnamefont {Sorel}}, \bibinfo {author}
  {\bibfnamefont {R.}~\bibnamefont {Morandotti}}, \bibinfo {author}
  {\bibfnamefont {N.}~\bibnamefont {Davidson}}, \ and\ \bibinfo {author}
  {\bibfnamefont {Y.}~\bibnamefont {Silberberg}},\ }\href {\doibase
  10.1103/PhysRevLett.103.013901} {\bibfield  {journal} {\bibinfo  {journal}
  {Phys. Rev. Lett.}\ }\textbf {\bibinfo {volume} {103}},\ \bibinfo {pages}
  {013901} (\bibinfo {year} {2009})}\BibitemShut {NoStop}%
\bibitem [{\citenamefont {Lahini}\ \emph {et~al.}(2010)\citenamefont {Lahini},
  \citenamefont {Bromberg}, \citenamefont {Christodoulides},\ and\
  \citenamefont {Silberberg}}]{Lahini_PRL10}%
  \BibitemOpen
  \bibfield  {author} {\bibinfo {author} {\bibfnamefont {Y.}~\bibnamefont
  {Lahini}}, \bibinfo {author} {\bibfnamefont {Y.}~\bibnamefont {Bromberg}},
  \bibinfo {author} {\bibfnamefont {D.~N.}\ \bibnamefont {Christodoulides}}, \
  and\ \bibinfo {author} {\bibfnamefont {Y.}~\bibnamefont {Silberberg}},\
  }\href {\doibase 10.1103/PhysRevLett.105.163905} {\bibfield  {journal}
  {\bibinfo  {journal} {Phys. Rev. Lett.}\ }\textbf {\bibinfo {volume} {105}},\
  \bibinfo {pages} {163905} (\bibinfo {year} {2010})}\BibitemShut {NoStop}%
\bibitem [{\citenamefont {Lahini}\ \emph {et~al.}(2011)\citenamefont {Lahini},
  \citenamefont {Bromberg}, \citenamefont {Shechtman}, \citenamefont {Szameit},
  \citenamefont {Christodoulides}, \citenamefont {Morandotti},\ and\
  \citenamefont {Silberberg}}]{Lahini_PRA11}%
  \BibitemOpen
  \bibfield  {author} {\bibinfo {author} {\bibfnamefont {Y.}~\bibnamefont
  {Lahini}}, \bibinfo {author} {\bibfnamefont {Y.}~\bibnamefont {Bromberg}},
  \bibinfo {author} {\bibfnamefont {Y.}~\bibnamefont {Shechtman}}, \bibinfo
  {author} {\bibfnamefont {A.}~\bibnamefont {Szameit}}, \bibinfo {author}
  {\bibfnamefont {D.~N.}\ \bibnamefont {Christodoulides}}, \bibinfo {author}
  {\bibfnamefont {R.}~\bibnamefont {Morandotti}}, \ and\ \bibinfo {author}
  {\bibfnamefont {Y.}~\bibnamefont {Silberberg}},\ }\href {\doibase
  10.1103/PhysRevA.84.041806} {\bibfield  {journal} {\bibinfo  {journal} {Phys.
  Rev. A}\ }\textbf {\bibinfo {volume} {84}},\ \bibinfo {pages} {041806}
  (\bibinfo {year} {2011})}\BibitemShut {NoStop}%
\bibitem [{\citenamefont {Guan}\ \emph {et~al.}(2012)\citenamefont {Guan},
  \citenamefont {Katz}, \citenamefont {Small}, \citenamefont {Zhou},\ and\
  \citenamefont {Silberberg}}]{Guan_OL12}%
  \BibitemOpen
  \bibfield  {author} {\bibinfo {author} {\bibfnamefont {Y.}~\bibnamefont
  {Guan}}, \bibinfo {author} {\bibfnamefont {O.}~\bibnamefont {Katz}}, \bibinfo
  {author} {\bibfnamefont {E.}~\bibnamefont {Small}}, \bibinfo {author}
  {\bibfnamefont {J.}~\bibnamefont {Zhou}}, \ and\ \bibinfo {author}
  {\bibfnamefont {Y.}~\bibnamefont {Silberberg}},\ }\href {\doibase
  10.1364/OL.37.004663} {\bibfield  {journal} {\bibinfo  {journal} {Opt.
  Lett.}\ }\textbf {\bibinfo {volume} {37}},\ \bibinfo {pages} {4663} (\bibinfo
  {year} {2012})}\BibitemShut {NoStop}%
\bibitem [{\citenamefont {Poem}\ \emph {et~al.}(2012)\citenamefont {Poem},
  \citenamefont {Gilead},\ and\ \citenamefont {Silberberg}}]{Poem_PRL12}%
  \BibitemOpen
  \bibfield  {author} {\bibinfo {author} {\bibfnamefont {E.}~\bibnamefont
  {Poem}}, \bibinfo {author} {\bibfnamefont {Y.}~\bibnamefont {Gilead}}, \ and\
  \bibinfo {author} {\bibfnamefont {Y.}~\bibnamefont {Silberberg}},\ }\href
  {\doibase 10.1103/PhysRevLett.108.153602} {\bibfield  {journal} {\bibinfo
  {journal} {Phys. Rev. Lett.}\ }\textbf {\bibinfo {volume} {108}},\ \bibinfo
  {pages} {153602} (\bibinfo {year} {2012})}\BibitemShut {NoStop}%
\bibitem [{\citenamefont {Verbin}\ \emph {et~al.}(2013)\citenamefont {Verbin},
  \citenamefont {Zilberberg}, \citenamefont {Kraus}, \citenamefont {Lahini},\
  and\ \citenamefont {Silberberg}}]{Verbin_PRL13}%
  \BibitemOpen
  \bibfield  {author} {\bibinfo {author} {\bibfnamefont {M.}~\bibnamefont
  {Verbin}}, \bibinfo {author} {\bibfnamefont {O.}~\bibnamefont {Zilberberg}},
  \bibinfo {author} {\bibfnamefont {Y.~E.}\ \bibnamefont {Kraus}}, \bibinfo
  {author} {\bibfnamefont {Y.}~\bibnamefont {Lahini}}, \ and\ \bibinfo {author}
  {\bibfnamefont {Y.}~\bibnamefont {Silberberg}},\ }\href {\doibase
  10.1103/PhysRevLett.110.076403} {\bibfield  {journal} {\bibinfo  {journal}
  {Phys. Rev. Lett.}\ }\textbf {\bibinfo {volume} {110}},\ \bibinfo {pages}
  {076403} (\bibinfo {year} {2013})}\BibitemShut {NoStop}%
\bibitem [{\citenamefont {Gilead}\ \emph {et~al.}(2015)\citenamefont {Gilead},
  \citenamefont {Verbin},\ and\ \citenamefont {Silberberg}}]{Gilead_PRL15}%
  \BibitemOpen
  \bibfield  {author} {\bibinfo {author} {\bibfnamefont {Y.}~\bibnamefont
  {Gilead}}, \bibinfo {author} {\bibfnamefont {M.}~\bibnamefont {Verbin}}, \
  and\ \bibinfo {author} {\bibfnamefont {Y.}~\bibnamefont {Silberberg}},\
  }\href {\doibase 10.1103/PhysRevLett.115.133602} {\bibfield  {journal}
  {\bibinfo  {journal} {Phys. Rev. Lett.}\ }\textbf {\bibinfo {volume} {115}},\
  \bibinfo {pages} {133602} (\bibinfo {year} {2015})}\BibitemShut {NoStop}%
\bibitem [{\citenamefont {Bai}\ \emph {et~al.}(2016)\citenamefont {Bai},
  \citenamefont {Xu}, \citenamefont {Lu}, \citenamefont {Zhong},\ and\
  \citenamefont {Zhu}}]{Bai_JO16}%
  \BibitemOpen
  \bibfield  {author} {\bibinfo {author} {\bibfnamefont {Y.~F.}\ \bibnamefont
  {Bai}}, \bibinfo {author} {\bibfnamefont {P.}~\bibnamefont {Xu}}, \bibinfo
  {author} {\bibfnamefont {L.~L.}\ \bibnamefont {Lu}}, \bibinfo {author}
  {\bibfnamefont {M.~L.}\ \bibnamefont {Zhong}}, \ and\ \bibinfo {author}
  {\bibfnamefont {S.~N.}\ \bibnamefont {Zhu}},\ }\href {\doibase
  10.1088/2040-8978/18/5/055201} {\bibfield  {journal} {\bibinfo  {journal} {J.
  Opt.}\ }\textbf {\bibinfo {volume} {18}},\ \bibinfo {pages} {055201}
  (\bibinfo {year} {2016})}\BibitemShut {NoStop}%
\bibitem [{\citenamefont {Avelar}\ \emph {et~al.}(2004)\citenamefont {Avelar},
  \citenamefont {Baseia},\ and\ \citenamefont {de~Almeida}}]{Avelar_JOB04}%
  \BibitemOpen
  \bibfield  {author} {\bibinfo {author} {\bibfnamefont {A.~T.}\ \bibnamefont
  {Avelar}}, \bibinfo {author} {\bibfnamefont {B.}~\bibnamefont {Baseia}}, \
  and\ \bibinfo {author} {\bibfnamefont {N.~G.}\ \bibnamefont {de~Almeida}},\
  }\href {\doibase 10.1088/1464-4266/6/1/007} {\bibfield  {journal} {\bibinfo
  {journal} {J. Opt. B Quantum Semiclassical Opt.}\ }\textbf {\bibinfo {volume}
  {6}},\ \bibinfo {pages} {41} (\bibinfo {year} {2004})}\BibitemShut {NoStop}%
\bibitem [{\citenamefont {Valverde}\ \emph {et~al.}(2003)\citenamefont
  {Valverde}, \citenamefont {Avelar}, \citenamefont {Baseia},\ and\
  \citenamefont {Malbouisson}}]{Valverde_PLA03}%
  \BibitemOpen
  \bibfield  {author} {\bibinfo {author} {\bibfnamefont {C.}~\bibnamefont
  {Valverde}}, \bibinfo {author} {\bibfnamefont {A.~T.}\ \bibnamefont
  {Avelar}}, \bibinfo {author} {\bibfnamefont {B.}~\bibnamefont {Baseia}}, \
  and\ \bibinfo {author} {\bibfnamefont {J.~M.~C.}\ \bibnamefont
  {Malbouisson}},\ }\href {\doibase 10.1016/S0375-9601(03)01049-1} {\bibfield
  {journal} {\bibinfo  {journal} {Phys. Lett. A}\ }\textbf {\bibinfo {volume}
  {315}},\ \bibinfo {pages} {213} (\bibinfo {year} {2003})}\BibitemShut
  {NoStop}%
\bibitem [{\citenamefont {Souza}\ \emph {et~al.}(2004)\citenamefont {Souza},
  \citenamefont {Avelar}, \citenamefont {de~Almeida},\ and\ \citenamefont
  {Baseia}}]{Souza_OC04}%
  \BibitemOpen
  \bibfield  {author} {\bibinfo {author} {\bibfnamefont {S.}~\bibnamefont
  {Souza}}, \bibinfo {author} {\bibfnamefont {A.~T.}\ \bibnamefont {Avelar}},
  \bibinfo {author} {\bibfnamefont {N.~G.}\ \bibnamefont {de~Almeida}}, \ and\
  \bibinfo {author} {\bibfnamefont {B.}~\bibnamefont {Baseia}},\ }\href
  {\doibase 10.1016/j.optcom.2004.05.039} {\bibfield  {journal} {\bibinfo
  {journal} {Opt. Commun.}\ }\textbf {\bibinfo {volume} {239}},\ \bibinfo
  {pages} {359} (\bibinfo {year} {2004})}\BibitemShut {NoStop}%
\bibitem [{\citenamefont {Walls}\ and\ \citenamefont
  {Milburn}(2008)}]{Walls_08}%
  \BibitemOpen
  \bibfield  {author} {\bibinfo {author} {\bibfnamefont {D.}~\bibnamefont
  {Walls}}\ and\ \bibinfo {author} {\bibfnamefont {G.~J.}\ \bibnamefont
  {Milburn}},\ }\href {\doibase 10.1007/978-3-540-28574-8} {\emph {\bibinfo
  {title} {{Quantum Optics}}}}\ (\bibinfo  {publisher} {Springer Berlin
  Heidelberg},\ \bibinfo {address} {Berlin, Heidelberg},\ \bibinfo {year}
  {2008})\BibitemShut {NoStop}%
\bibitem [{\citenamefont {Christodoulides}\ \emph {et~al.}(2003)\citenamefont
  {Christodoulides}, \citenamefont {Lederer},\ and\ \citenamefont
  {Silberberg}}]{Christodoulides_NAT03}%
  \BibitemOpen
  \bibfield  {author} {\bibinfo {author} {\bibfnamefont {D.~N.}\ \bibnamefont
  {Christodoulides}}, \bibinfo {author} {\bibfnamefont {F.}~\bibnamefont
  {Lederer}}, \ and\ \bibinfo {author} {\bibfnamefont {Y.}~\bibnamefont
  {Silberberg}},\ }\href {\doibase 10.1038/nature01936} {\bibfield  {journal}
  {\bibinfo  {journal} {Nature}\ }\textbf {\bibinfo {volume} {424}},\ \bibinfo
  {pages} {817} (\bibinfo {year} {2003})}\BibitemShut {NoStop}%
\bibitem [{\citenamefont {Eisenberg}\ \emph {et~al.}(1998)\citenamefont
  {Eisenberg}, \citenamefont {Silberberg}, \citenamefont {Morandotti},
  \citenamefont {Boyd},\ and\ \citenamefont {Aitchison}}]{Eisenberg_PRL98}%
  \BibitemOpen
  \bibfield  {author} {\bibinfo {author} {\bibfnamefont {H.~S.}\ \bibnamefont
  {Eisenberg}}, \bibinfo {author} {\bibfnamefont {Y.}~\bibnamefont
  {Silberberg}}, \bibinfo {author} {\bibfnamefont {R.}~\bibnamefont
  {Morandotti}}, \bibinfo {author} {\bibfnamefont {A.~R.}\ \bibnamefont
  {Boyd}}, \ and\ \bibinfo {author} {\bibfnamefont {J.~S.}\ \bibnamefont
  {Aitchison}},\ }\href {\doibase 10.1103/PhysRevLett.81.3383} {\bibfield
  {journal} {\bibinfo  {journal} {Phys. Rev. Lett.}\ }\textbf {\bibinfo
  {volume} {81}},\ \bibinfo {pages} {3383} (\bibinfo {year}
  {1998})}\BibitemShut {NoStop}%
\bibitem [{\citenamefont {Bromberg}\ \emph {et~al.}(2009)\citenamefont
  {Bromberg}, \citenamefont {Lahini}, \citenamefont {Morandotti},\ and\
  \citenamefont {Silberberg}}]{Bromberg_PRL09}%
  \BibitemOpen
  \bibfield  {author} {\bibinfo {author} {\bibfnamefont {Y.}~\bibnamefont
  {Bromberg}}, \bibinfo {author} {\bibfnamefont {Y.}~\bibnamefont {Lahini}},
  \bibinfo {author} {\bibfnamefont {R.}~\bibnamefont {Morandotti}}, \ and\
  \bibinfo {author} {\bibfnamefont {Y.}~\bibnamefont {Silberberg}},\ }\href
  {\doibase 10.1103/PhysRevLett.102.253904} {\bibfield  {journal} {\bibinfo
  {journal} {Phys. Rev. Lett.}\ }\textbf {\bibinfo {volume} {102}},\ \bibinfo
  {pages} {253904} (\bibinfo {year} {2009})}\BibitemShut {NoStop}%
\bibitem [{\citenamefont {Crank}\ and\ \citenamefont
  {Nicolson}(1947)}]{Crank_MPCPS47}%
  \BibitemOpen
  \bibfield  {author} {\bibinfo {author} {\bibfnamefont {J.}~\bibnamefont
  {Crank}}\ and\ \bibinfo {author} {\bibfnamefont {P.}~\bibnamefont
  {Nicolson}},\ }\href {\doibase 10.1017/S0305004100023197} {\bibfield
  {journal} {\bibinfo  {journal} {Math. Proc. Cambridge Philos. Soc.}\ }\textbf
  {\bibinfo {volume} {43}},\ \bibinfo {pages} {50} (\bibinfo {year}
  {1947})}\BibitemShut {NoStop}%
\bibitem [{\citenamefont {Press}\ \emph {et~al.}(2002)\citenamefont {Press},
  \citenamefont {Teukolsky}, \citenamefont {Vetterling},\ and\ \citenamefont
  {Flannery}}]{Press}%
  \BibitemOpen
  \bibfield  {author} {\bibinfo {author} {\bibfnamefont {W.~H.}\ \bibnamefont
  {Press}}, \bibinfo {author} {\bibfnamefont {S.~A.}\ \bibnamefont
  {Teukolsky}}, \bibinfo {author} {\bibfnamefont {W.~T.}\ \bibnamefont
  {Vetterling}}, \ and\ \bibinfo {author} {\bibfnamefont {B.~P.}\ \bibnamefont
  {Flannery}},\ }\href@noop {} {\emph {\bibinfo {title} {{Numerical recipes in
  C++: The Art of Scientific Computing}}}},\ \bibinfo {edition} {2nd}\ ed.\
  (\bibinfo  {publisher} {Cambridge University Press},\ \bibinfo {address} {New
  York},\ \bibinfo {year} {2002})\BibitemShut {NoStop}%
\end{thebibliography}%

\end{document}